\documentclass[fleqn,usenatbib]{mnras}

\usepackage{newtxtext,newtxmath}

\usepackage[T1]{fontenc}

\DeclareRobustCommand{\VAN}[3]{#2}
\let\VANthebibliography\thebibliography
\def\thebibliography{\DeclareRobustCommand{\VAN}[3]{##3}\VANthebibliography}

\usepackage{graphicx}	
\usepackage{amsmath}	

\usepackage{amssymb}	
\usepackage{longtable}

\title[Assessing Residual Images Using Deep Learning]{Characterization of Residual Morphological Substructure Using Supervised and Unsupervised Deep Learning}

\author[K.B. Mantha et al.]{Kameswara Bharadwaj Mantha,$^{1}$\thanks{E-mail: km4n6@mail.umkc.edu}
Daniel H. McIntosh$^{1}$, Cody Ciaschi$^{1}$, Rubyet Evan$^{1}$, Luther Landry$^{1}$,
\newauthor Henry C. Ferguson$^{2}$, Camilla Pacifici$^{2}$, Joel Primack$^{3}$, Nimish Hathi$^{2}$, Anton Koekemoer$^{2}$, Yicheng Guo$^{4}$, 
\newauthor $+$ The CANDELS Collaboration.
\\
$^{1}$Department of Physics \& Astronomy, University of Missouri Kansas City, Kansas City, MO, 64110, USA.\\
$^{2}$ Space Telescope Science Institute, 3700 San Martin Drive, Baltimore, MD
21218, USA.\\
$^{3}$ Department of Physics, University of California, Santa Cruz, CA 95064, USA.\\
$^{4}$ Department of Physics and Astronomy, University of Missouri, Columbia, MO 65211, USA.}

\date{\bf This manuscript is a preprint that has not undergone peer review and is being shared to ensure  dissemination and community access to the results and insights (see acknowledgements).}

\pubyear{2021}

\begin{document}
\label{firstpage}
\pagerange{\pageref{firstpage}--\pageref{lastpage}}
\maketitle

\begin{abstract}
{Automated identification and characterization of galactic substructure is an essential step in understanding the transformative physical processes driving the galaxy evolution. In this study, we investigate the application of deep learning (DL) frameworks to characterize different kinds of galactic substructures hosted within parametric light-profile subtracted ``residual'' images of a large sample galaxies from the CANDELS survey. We develop a supervised Convolutional Neural Network (CNN) and unsupervised Convolutional Variational Autoencoder (CvAE) and train it on the single-S\'ersic profile fitting based residual images of $10,046$ bright and massive galaxies ($H<24.5\,{\rm mag}$ and $M_{\rm stellar} \geq 10^{9.5}\,M_{\odot}$) spanning $1<z<3$, in conjunction with their visual-based classification labels (five classes; for CNN) indicating the nature of residual substructures hosted within them. Using our unique data pre-processing approach, we prepare our residual image data such that the input images to our DL networks only comprise the ``galaxy of interest'', and augment the data set such that our training sample spans uniformly across different residual characteristics and relative viewing angles. We assess the latent space of the CNN and CvAE using Principle Component Analysis (PCA) along with independently quantified metrics of residual strength (significant pixel flux -- $SPF$, Bumpiness -- $B$, and Residual Flux Fraction -- $RFF$). We also employ an unsupervised Gaussian Mixture Modeling (GMM) based clustering scheme with Support Vector Classification (SVC) to identify groupings (and their decision boundaries) in PCA space that correspond to similar residual substructure. We find that our supervised CNN latent features in PCA space correlate with the $SPF$ values and distinguish between qualitatively strong and weak residual substructures. While our unsupervised CvAE latent space also correlates with visual and quantitative residual characteristics, but lacks clear discriminatory power (compared to supervised CNN) when characterizing different residual substructures.}
\end{abstract}

\begin{keywords}
galaxy substructure -- deep learning -- representation learning
\end{keywords}



\section{Introduction}
{Major galaxy merging (mass ratio $<4$) is a fundamental aspect of the hierarchical structure-growth scenario of the Universe. As such, it is theoretically predicted to contribute to the empirically well-documented stellar-mass growth of high-mass galaxies (e.g., $M_{\rm stellar}>10^{11}M_{\odot}$), enhancing global galactic star-formation, and facilitating central super-massive black hole growth and triggering. Observationally identifying major mergers is a key methodological step in empirically verifying the ``merging -- galaxy evolution'' connection.  Commonly, {\it close-pair} and {\it morphological} methods are used to identify major mergers, among which quantitative metrics (e.g., $CAS$, $G-M_{20}$ parameters) and qualitative visual classification of galaxy images or light-profile subtracted residual images (e.g., from {\tt GALFIT}) are frequently-employed approaches to assess the presence of disturbed morphological substructure. While these methods yield broadly-consistent, rising merger frequency evolution measurements out to $z\sim 2$, each approach selects from a distribution of merger stages (e.g., pre-coalescence vs. late-stage), and suffers from systematic selection biases caused by cosmological surface-brightness dimming, different observability timescales, and subjective interpretation of low surface-brightness signatures. Advancements in machine learning has enabled recent efforts to overcome some of these biases through supervised frameworks and unsupervised deep learning (DL) networks to better identify mergers out to $z\sim 3$ and accelerate visual classifications. In this study, we develop two new public unsupervised and supervised DL frameworks and demonstrate their applicability on {\tt GALFIT}-based residual images of $\sim 10,000$ massive galaxies spanning $1<z<3$ to perform automated residual substructure characterization and accelerate visual classification of especially interpretive classes (e.g., plausible hallmark tidal features) that impose subjective assessment of physical processes. The machine learning tools developed in this work are timely in the era of big-data astrophysics and applicability to large-scale surveys by upcoming telescope facilities. }

Galaxy-galaxy mergers are theoretically predicted to happen as a natural consequence of the hierarchical, gravitation-driven merging of dark-matter halos. Numerical simulations predict galaxy merging as an important pathway for several key aspects of galaxy evolution such as stellar-mass growth and buildup of massive galaxies \citep{Springel00,Naab06,Cox08}, enhancement of star-formation \citep{Sanders88a,Dimatteo07,DiMatteo08}, growth and triggering of central super-massive black hole \citep{Hopkins06a,Younger09,Narayanan10,hopkins_mergers_2010}. Indeed, many observational studies on these fronts have found evidence for a strong ``merger-galaxy evolution'' connection supporting the theoretical predictions \citep{bell06b,mcintosh_2008,van_der_wel09,Treister12,weston17}.  However, some studies have also found contradictory results that contest the hypothesized strong role of mergers in galaxy evolution \citep{robaina09,Swinbank10,kocevski_candels:_2012,villforth17}, hinting at alternate physical processes that may be at simultaneous play \citep[e.g., Violent Disk Instabilities and Cold-flow accretion;][]{Bournaud11,Dekel14,Ceverino15a}. This highlights that the role of major mergers in galaxy evolution is a key open question, and robustly identifying major merging systems and quantifying their incidence is a key methodological step.

Motivated by the long-standing numerical expectation based idea that galaxies in close physical proximity will interact gravitationally and merge into a more-massive system, several studies over the past two decades have used {\it close-proximity pairs} as probes for ongoing or future merging systems and quantified their incidence over a wide redshift range $0\lesssim z\lesssim 6$. These broadly agree that close pair frequency rises between the redshifts $0\lesssim z\lesssim 1.5$, albeit with a wide range of redshift dependencies $\sim (1+z)^{0.5-3}$ \citep{Zepf89,Patton97,Lin04,de_ravel09,de_ravel_zcosmos_2011,Robotham14,man16,Mundy17,Mantha18,Duncan19}, owing to the differences in employed pair selection choices. \cite{Mantha18} analyzed the effects of different close-pair selection choices on the derived pair fractions. However, despite using closely matched pair selection criteria, recent measurements of $f_{\rm pair}(z)$ evolution at $z>1.5$ vary from moderately increasing to flat, or diminishing trends \citep{williams_diminishing_2011,man16,Mundy17,Ventou17,Mantha18,Duncan19}. This may be due to important study-to-study variant observational effects such as photometric redshifts and their errors, stellar-mass uncertainties, sample incompleteness, and intrinsic field-to-field pair fraction variations. These aspects are being investigated in an ongoing study and will be discussed in an upcoming paper (Mantha et al., {\it in prep}).

Simultaneously, several studies have also used {\it morphological methods} to identify merging systems, motivated by the theoretical merger simulation predictions that merging galaxies exhibit morphological disturbances induced by gravitational tidal forces during the interaction. Most morphology-based studies often adopt subjective visual identification of disturbed morphology \citep{Wolf05,bell_dry_2006,Jogee09} or quantitative metrics of large-scale galaxy morphology such as Gini-M$_{20}$ \citep{Abraham03,lotz08,conselice_structures_2009}, $CAS$ \citep{Abraham96a,Abraham96b,conselice_03_cas,conselice_structures_2008,lopez-sanjuan_galaxy_2009} parameters. These morphology based studies find a wide range of (often strongly evolving)  merger frequency evolutionary trends $\propto (1+z)^{2-5}$ at $z\lesssim 1.5$ \citep{lotz_major_2011}, albeit with some finding no redshift dependence \citep{cassata05,lotz08}, and some noting $\sim 25-50\%$ merger fractions at $z\sim 2-3$ \citep{conselice03a}. Such variations are due to morphological-study specific systematic effects such as small number statistics owing to narrow pencil-beam surveys, redshift dependent cosmological surface brightness dimming, morphological {\it k}-corrections when using bluer rest-frame wavelengths, that can impact the completeness and purity of the selected merger samples. Some recent studies have used latest {\it Hubble Space Telescope} ({\it HST}) based data \citep[e.g., CANDELS survey][]{Koekemoer,grogin_candels:_2011} in conjunction with more focused indicators of mergers (e.g., multiple nuclei in close proximity, faint and transient tidal features; \citep{Lackner14,Mantha19}) as a means to evade from some of such systematics. 

Tidal features are theoretically predicted to be prevalent among merging galaxies \citep{Toomre72,Eneev73,BarnesHernquist96,Duc2013}, and previous studies have identified galaxies hosting merging activity by detecting tidal features using qualitative visual inspections \citep{bell_dry_2006,mcintosh_2008} and quantitative residual metrics \citep{Tal_14,Hoyos12}  of the residual images produced by parametric light-profile fitting software  \citep[e.g., GALFIT;][]{Peng02}. However, these studies often rely on the subjective visual interpretations of human classifiers that can be time intensive and non-repeatable (especially for large-scale surveys), or use quantitative metrics that only indicate the plausibility of tidal signatures and do not quantitatively capture their key properties (strength, shape, color, etc). In an attempt to overcome some of these pitfalls, \citep{Mantha19}, we developed a more direct quantitative way to extract and quantify the strengths of residual substructure, including tidal features hosted by CANDELS galaxies. However, the sample used to demonstrate our feature extraction tool is still based on select galaxies identified through visual inspection. This motivates the strong need to develop automated methods for characterizing residual substructures and providing an accelerated path towards performing a more focused visual/quantitative assessment of galaxies hosting highly interpretive signatures (e.g., tidal features). One methodological avenue that has seen recent booming advancements in the field of image-based automated characterization is {\it Deep Learning}.

Deep learning (DL) conceptually allows the user to automatically extract low-level pixel features (often referred to as latent space) by iteratively adjusting a series of tuneable parameters within the ``deep'' layers such that the DL network attempts to achieve an overall global (user-defined) goal. This automated extraction of latent space can be broadly categorized into these approaches -- user-guided (i.e., supervised) learning, unguided (unsupervised) learning, and sparsely-guided (semi-supervised) learning. The extracted latent space can be used for further regression/classification tasks or for dimensionality reduction purposes. Many recent studies have adopted DL frameworks for various astronomical applications such as galaxy morphology classifications \citep{Huertas-Company15,Dieleman15,Dominguez-Sanchez18,Huertas-Company18}, gravitational lens finding \citep{Hocking18,Metcalf19,Cheng20}, galaxy merger identifications \citep{Ferreira20,Tohill21}. Several of these morphological or merger related  DL-based studies often train Convolutional Nueral Networks \citep[CNNs; for supervised learning;][]{Fukushima82,Lecun98} or Convolutional Autoencoders \citep[CAE; for unsupervised learning;][]{Masci11} on the single or multi-wavelength band images of galaxies from the telescopic imaging. However, such an approach may not be ideal in the case of galaxy merger identifications where their signatures are faint and are sub-dominant when compared to the global galactic light profile.  As such, devising DL networks to train on the residual images (after subtracting the galactic light profile) may be much more fruitful, yet this approach is relatively unexplored. Only recently, \cite{Storey-Fisher21} developed Generative Adversarial Networks to model multi-band galaxy images and used the resultant model-subtracted images to identify ``anomalous'' galaxies. Moreover, it is a common practice among large astronomy survey science communities to run light-profile fitting software  \citep[e.g., GALFIT;][]{Peng02} and generate residual images for a large galaxy samples. Therefore, analyzing residual images using DL frameworks remains a worthwhile exploratory path, which can provide provide insightful information and automated assessment of the morphological substructure hosted by large galaxy samples.

In this DL-based exploratory study, we develop supervised CNN and unsupervised variational CAE (CvAE) networks to characterize different kinds of residual substructures hosted by a large sample of $\sim 10,000$ massive and bright galaxies from the {\it HST} CANDELS survey spanning $1<z<3$. Our analysis focuses primarily on these networks' deep latent space extraction and we use Principle Component Analysis (PCA) to assess the extracted latent space of our galaxies using qualitative (i.e., visual based labels) and quantitative metrics. We also explore unsupervised clustering in dimensionality-reduced PCA space and assess the supervised and unsupervised DL networks' ability to automatically distinguish different residual substructures at a latent space level.  

We structure this paper as follows -- In \S\,\ref{sec:candels_data}, we describe our main CANDELS massive and bright galaxy sample selection, its corresponding imaging data products, and visual characteristic catalogs of residual images used for training and assessment of our DL models. In \S\,\ref{sec:preprocessing}, we describe our training and testing sample construction, data augmentation steps, imaging data preparatory steps involving creation of scaled ``object-only'' residual images that serve as inputs to our DL frameworks, and computation of independent residual quantitative metrics used for latent space assessments. In \S\,\ref{sec:models}, we introduce our supervised and unsupervised architectures and discuss their training and optimization strategy. In \S\,\ref{sec:results}, we analyze the deep latent space from our supervised and unsupervised frameworks using PCA, followed by unsupervised clustering in PCA space, and assess of our networks' abilities to characterize different residual substructure features. Finally, we present our concluding statements in \S\,\ref{sec:conclusions}.

\section{CANDELS Data}\label{sec:candels_data}
{To implement our supervised and unsupervised DL frameworks for an automated residual substructure characterization, we use {\it HST}/WFC3 {\it H}-band (F160W) images and {\it GALFIT}-based single-S\'ersic model-subtracted residual images by \cite{van_der_wel_12} for a sample of $10,046$ massive galaxies ($M_{\rm stellar}\geq 10^{9.5}M_{\odot}$) spanning the key epoch of galaxy evolution $1<z<3$. { For training our supervised model and independently assessing the learning of our unsupervised network, we also use residual characteristic flags based on visual inspection from an ongoing effort ({\it HST}-AR 15040; PI: McIntosh). }}

\subsection{Source Catalogs and Sample Selection}
{To select our main galaxy sample used in this work, we use the source catalogs and high-level science data products from the CANDELS survey \citep{Koekemoer,grogin_candels:_2011}, which spans a total sky area of $\sim 800\,{\rm arcmin^{2}}$ over five {\it HST} legacy fields -- UDS, GOODS-S, GOODS-N, COSMOS, and EGS. We use the standard source extraction \cite{bertin96} based photometric source catalogs generated using the {\it HST} {\it H}-band ($2$-orbit) imaging by \citep{galametz_candels_2013,guo_candels_2013,Barro19,Nayyeri17,Stefanon17} for the five fields, respectively. Each identified source in these catalogs are provided with a quality flag (PhotFlag) following a robust automated routine by \citep{galametz_candels_2013} to identify contaminated sources due to their proximity to nearby stars or image edges. We use a PhotFlag$=0$ to exclude such contaminant sources ($\sim 3-5\%$ of the overall identified sources). To ensure that our sources have robust structural parameter fits and reliable visual characterization of residual substructure (as discussed later in this section), we enforce a {\it H}-band magnitude cut and select sources that are brighter than $H=24.5\,{\rm mag}$.}

For these {\it H}-band bright sources, we then use the latest best-available redshift ($z_{\rm best}$) information by \cite{Kodra19}, who used bayesian methods to combine the photometric redshift probability distributions from several participants in a previous CANDELS-team effort \citep{dahlen_critical_2013}, and provided latest photometric redshifts and their uncertainties while giving precedence to the relatively more precise {\it HST} Grism or spectroscopic redshifts wherever available. Additionally, we also use the CANDELS-team stellar-mass estimates ($M_{\rm stellar}$) generated by \cite{santini_stellar_2015,Mobasher15}, who performed SED-fitting of multi-band photometric data spanning optical to near-IR wavelengths over a range of template-fitting codes and physical parameter assumptions \citep[see][]{Mobasher15}. Using the redshift ($z_{\rm best}$) and stellar-mass ($M_{\rm stellar}$) information, we select our main galaxy sample of $10,046$ massive galaxies with $M_{\rm stellar}\geq10^{9.5}\,M_{\odot}$ and $H<24.5\,{\rm mag}$, and spanning a redshift range $1\leq z_{\rm best}\leq 3$.

\subsection{Imaging Products and {\tt GALFIT} Residuals}
\begin{figure}
    \centering
    \includegraphics[width=0.95\columnwidth]{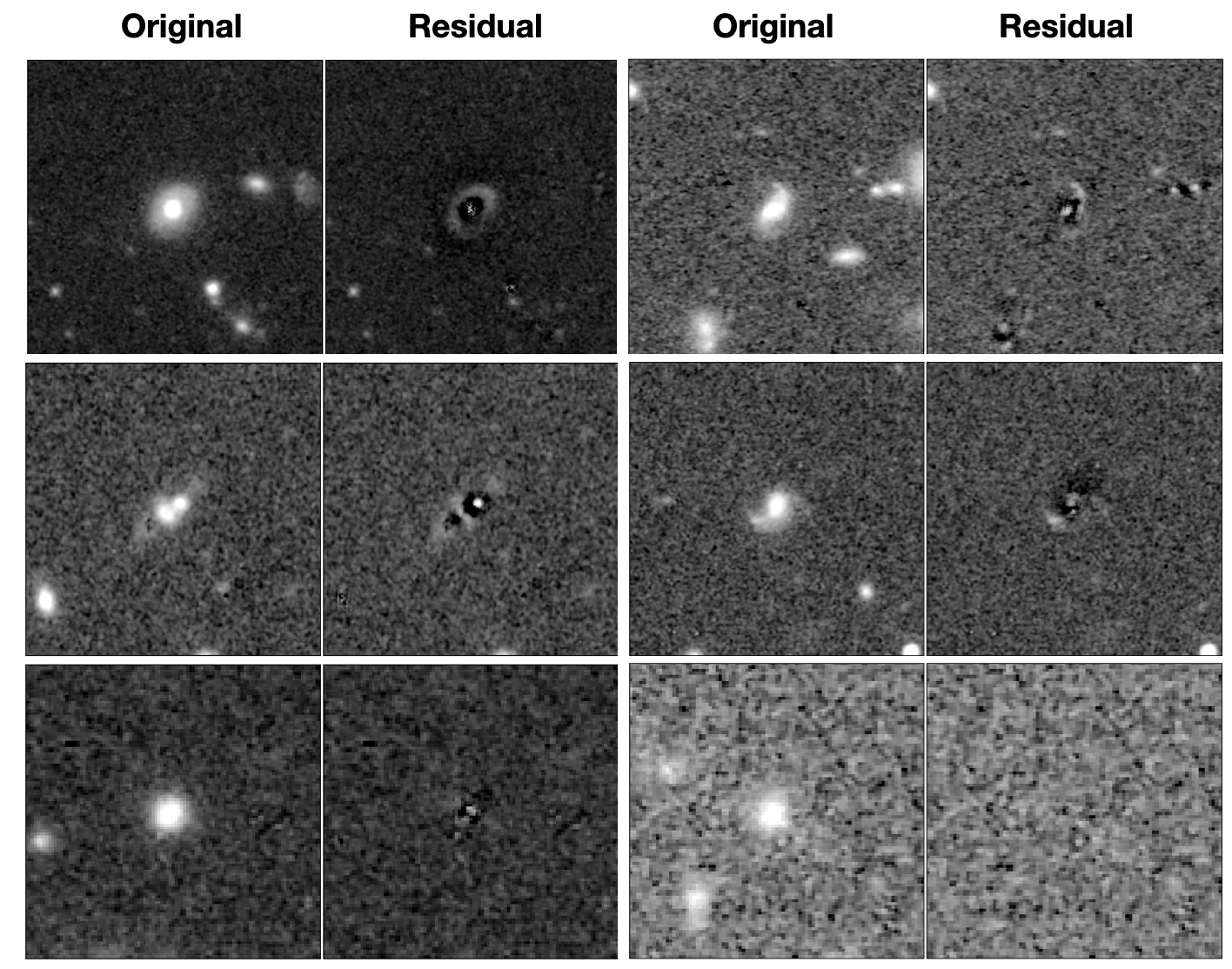}
    \caption[Example visualization of original and residual images]{Visualization of the {\it H}-band original images (left in each column pair) and their corresponding residual images (right panel in each column pair) for example CANDELS galaxies from our main sample, generated by performing single-S\'ersic light-profile fitting using {\tt GALFIT} \citep{Peng02} by \cite{van_der_wel_12}.}
    \label{fig:origina_and_residual_visualization}
\end{figure}

{We use the {\it H}-band structural fitting data by \cite{van_der_wel_12}, who used a popular light-profile fitting software {\tt GALFIT} \citep{Peng02} to model the galaxy's structural parameters identified in the CANDELS source catalogs. Briefly, \cite{van_der_wel_12} employed the standard procedures outlined in \cite{Barden12} and used the CANDELS {\it H}-band image mosaics \citep{Koekemoer} in conjunction with a comprehensive SExtractor \citep{bertin96} routine to identify sources, make postage stamp cutouts, mask unwanted sources\footnote{\cite{van_der_wel_12} masked neighbouring sources $4\,{\rm mag}$ fainter than the primary object, and all remaining sources were simultaneously fit.}, compute initial structural parameter guesses, compute a robust sky-background, and fit each source with a single S\'ersic profile. For each CANDELS galaxy, \cite{van_der_wel_12} generated an image-cube using {\tt GALFIT} \citep{Peng02}, which contains the {\it original} image used for fitting, the {\it model} image with the best-fit structural parameter information stored in its header, and the {\it residual} (original$-$model) image. In Figure\,\ref{fig:origina_and_residual_visualization}, we show example {\it H}-band original images of our galaxies along with their corresponding residuals by \cite{van_der_wel_12}. In a later section (\S\,\ref{sec:preprocessing}), we further process each galaxy's residual images to prepare them for our DL-based analysis.  }

\subsection{Visual { Inspection} of Residual Substructure}\label{sec:residual_characterization}
\begin{figure*}
    \centering
    \includegraphics[width=1.9\columnwidth]{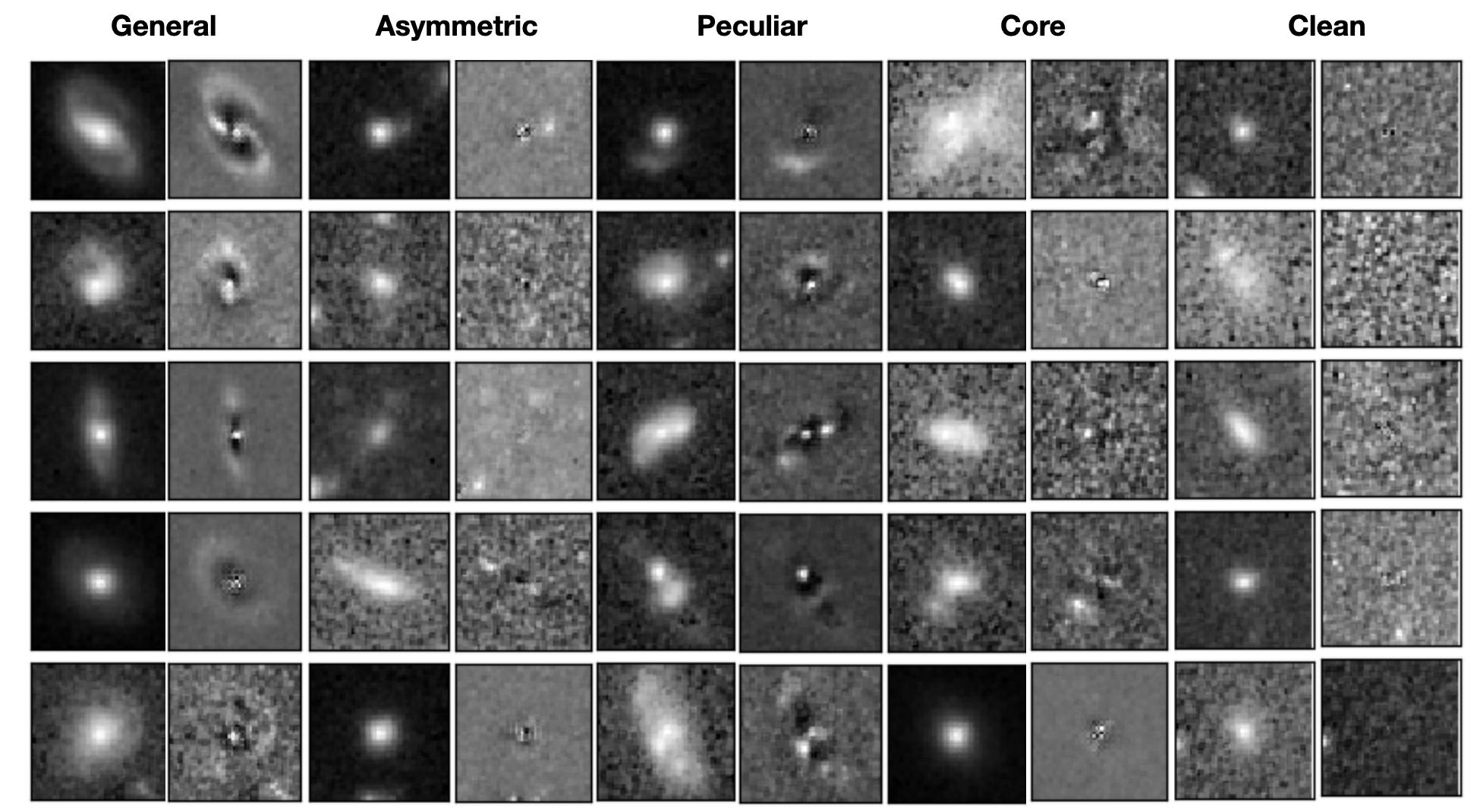}
    \caption[Example visualization of different residual substructures hosted by CANDELS galaxies]{An example view of our galaxies and their hosted residual substructure. In each column pairs, we show {\it HST} WFC3/F160W {\it H}-band {original} images (left) and their corresponding  {\tt GALFIT}-based residual images from \cite{van_der_wel_12} (right) for five example galaxies per class.}
    \label{fig:residual_visualization}
\end{figure*}

{To guide the training of our supervised DL network with user-defined labels that describe the residual substructure and qualitatively assess the outcomes of our unsupervised approach, we construct a catalog of visually-identified residual characteristics hosted for our main galaxy sample. { Briefly, we visually inspected all the residual images, following a process in which} at least three human classifiers simultaneously inspected each galaxy's original and residual images to characterize the visual nature of the hosted residual substructure. This information is validated by an expert and is further processed using a $2/3^{\rm rd}$ majority voting to generate a catalog of residual characteristics for our galaxy sample. We broadly bin them into five classes that span a range of residual feature characteristics and qualitative strengths, and an additional quality check flag indicating plausible fit-quality issues. { In Table\,\ref{tab:visual_char_table}, we show the definitions used while visually inspecting our sample residual images and we report the percentage breakdown of different residual characteristic groupings. Briefly, the ``Clean'' residuals have no significant left-over light and are consistent with a background noise appearance. The ``General'' residuals have strong left-over light, usually corresponding to a disk or spiral shape, or clumpy substructure associated with the overall disk or spiral morphology of the galaxy. Peculiar residuals also have strong under-fit residuals, however, they are usually asymmetric and align with the overall disturbed morphological structure of the galaxy. The ``Core'' residuals have a bright point-like under-fit structure towards the central region of the galaxy. Finally, ``Asymmetric'' residuals have concentrated or diffuse non-symmetric signatures with or without the presence of central excess.}

\begin{table*}
    \centering
    \caption{Breakdown of the five residual characteristic classes based on the visual inspection of our $10,046$ massive galaxy sample from the CANDELS survey. Columns: (1) Name of the residual characteristic; (2) Adopted definition of the residual characteristic for the visual inspection; (3) Number of galaxies in the sample with the corresponding residual characteristic; (4) Percentage contribution of the residual characteristic out of $10,046$ galaxies. }
    \label{tab:visual_char_table}
    \begin{tabular}{p{3cm}|p{6.5cm}|p{2cm}|p{2cm}} 
    	\hline
    	Residual Characteristic & Definition   &   Sample Size & Percentage \\
    	(1) & (2)   &  (3) & (4) \\
    	\hline
    	{\bf Clean} & The residual has no apparent features or structures pertaining to the galaxy and is consistent with the background noise. & $3815$ & $38.0\%$ \\
    	\hline
    	{\bf General} & The residual hosts strong left-over light with either a symmetric disk or spiral-like structure, or clumpy under-fit substructure associated to underlying plausible disk or spiral galaxy morphology. & $608$ & $6.0\%$ \\
    	\hline
    	{\bf Core} & The residual contains one or more bright point-like grouping of pixels near to the central region (by eye) of the galaxy, and no excess away from the galaxy's central region. & $1958$ & $19.5\%$\\
    	\hline
    	{\bf Asymmetric} & The residual contains concentrated or diffuse, asymmetric excess light outside the central region of the galaxy, with or without the presence of central excess light. & $3043$ & $30.3\%$ \\
    	\hline
    	{\bf Peculiar} & The residual contains strong asymmetric excess light with or without multiple central point excess that aligns with the overall disturbed morphology of the galaxy. & $512$ & $5.1\%$ \\
    	\hline
    	{\bf Fit-Quality Issue} & The galaxy's residual region of interest is very close to the image edges or artefacts, or is dominated/contaminated by a nearby star's light. & $110$ & $1.1\%$ \\
    	\hline
    \end{tabular}

\end{table*}

{ We note that a dominant portion of our main sample are qualitatively Clean and Asymmetric residuals ($\sim 38\%$ and $\sim 30\%$, respectively), and $\sim 19\%$ of our sample hosts cored residuals (see Table\,\ref{tab:visual_char_table}). A sub-dominant, but notable fraction of our sample ($\sim 11\%$) are within the General and Peculiar classes, hosting qualitatively strong residual characteristics.} In Figure\,\ref{fig:residual_visualization}, we show example {\it original} and {\it residual} images of our galaxies per residual characteristic class, where different residual substructures as per the class-wise definitions are visually evident.  We note that the visual inspection based residual characteristics information used in this work is preliminary { and pending full validation by an expert}. { While a future fully-validated sample could change the results discussed here, our current analysis and methodological framework are meant to serve as an informative exercise to aid the full validation of our residual characterization exercise.} }

\section{Data Preparation}\label{sec:preprocessing}
{Thus far, we have discussed the selection of our main galaxy sample and their imaging and cataloged data products. To prepare this data for our DL analysis, we carryout the following key pre-processing steps -- i) generation of residual image cutouts which only contain information corresponding to the ``galaxy-of-interest'' (GOI); ii)  construction of training and testing sub-samples and generating data augmentation applied training sample; iii) computation of independent residual metrics to aid the assessment of the learning of our DL frameworks.}

\subsection{Creating Object-Only Image for Training and Testing Samples}
\begin{figure}
    \centering
    \includegraphics[width=1\columnwidth]{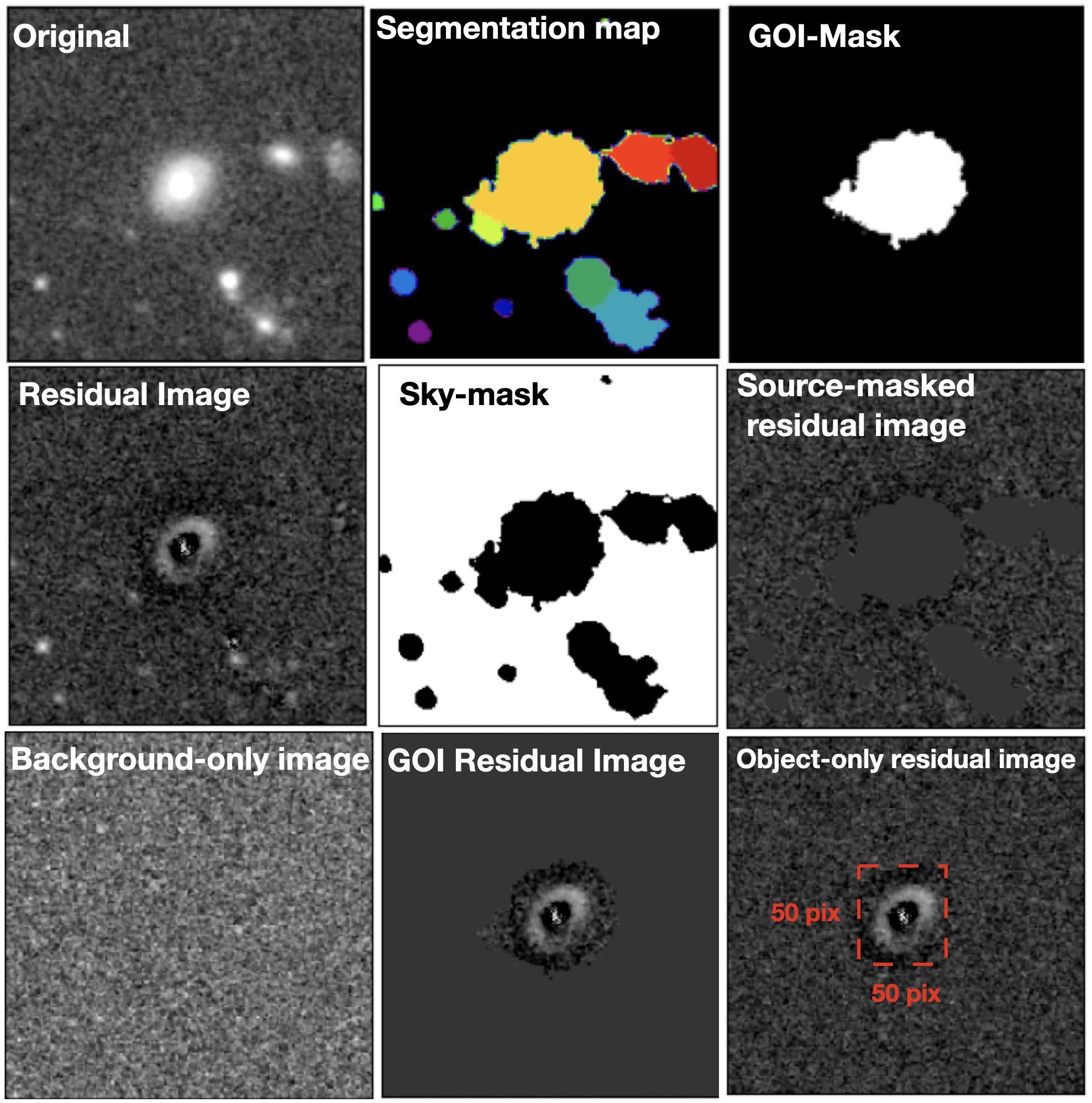}
    \caption[Step-wise visualization of image data preparation for deep-learning frameworks]{Visual illustration of our preprocessing step to prepare our residual images for the DL framework. We show the original image, source extraction based segmentation map highlighting the detected sources, residual image, region of interest corresponding the central galaxy in the image, and the GOI region of the residual image with a $50\,{\rm pix}\times 50\,{\rm pix}$ postage stamp size for reference.}
    \label{fig:preprocessing}
\end{figure}

{Each galaxy's image may contain other objects in it's field of view, which are not the main focus of the residual characteristics { for the GOI} (by design at the center of the image, see \cite{van_der_wel_12}). Furthermore, such interlopers may confuse the low-level, feature learning of the DL model. Therefore, we process the residual images such that only the pixel data corresponding to the GOI is preserved and the rest is matched with the sky background. 

In Figure\,\ref{fig:preprocessing}, we visually demonstrate the steps involved in the generation of object-only images for one example galaxy. We first identify sources using {\tt SEP} \citep{barbary16}, which is a python adaptation of the standard source extraction software SExtractor \citep{bertin96}. We adopt the source-extraction settings used in \cite{Mantha19}, where we use a 2D Tophat convolution kernel (radius$=5\,{\rm pix}$), a minimum source significance of $0.75$, and a minimum area of $7\,{\rm pix}$. { These adopted settings help identify small and faint sources}, and enable their segmentation regions to { encompass the light reaching far out from the galaxy's center}. { This enables robust quantification of sky background and masking of close projected-sky proximity interlopers} \citep[see][for more details]{Mantha19}. We then use the segmentation region map generated during the source-extraction step (top-center panel in Figure\,\ref{fig:preprocessing}) to generate a binary {GOI-mask} (top-right in Figure\,\ref{fig:preprocessing}), where the pixels corresponding to the GOI are set to one, and rest all other interlopers including the sky background are set to zero. Simultaneously, we also generate a generate a binary sky-mask (middle center panel in Figure\,\ref{fig:preprocessing}) by making all the source segmentation regions to zero, and the background to one. We then generate a source-masked, background sky-only image by multiplying the sky-mask with the residual image (center-right panel). We randomly sample $5\times5$ pixel regions from this background sky image to construct a contiguous background-sky image (bottom-left panel) matching the residual image size, where the native small-scale pixel-to-pixel correlation information is preserved. We then generate the GOI residual image (bottom-center panel) by multiplying the GOI-mask with the residual image data. Finally, we generate an ``object-only'' residual image (bottom-right panel) by adding the GOI residual image and the GOI-masked background-sky image (background-sky $\times$ GOI-mask$^{\rm c}$), where GOI-mask$^{\rm c}$ is the complement (inverted) image of the GOI-mask. 

{ Following the standard practice used in DL-based literature  \citep[e.g.,][]{Huertas-Company15,Metcalf19,Ferreira20}, we use the cutouts of the object-only images as inputs to our DL frameworks. As will be discussed in \S\,\ref{sec:models}, our DL-based frameworks are inspired by the shallow network architecture of \cite{Huertas-Company19}, and therefore we choose $50{\,\rm pix}\times 50\,{\rm pix}$ as our cutout dimensions centered at the {\tt GALFIT}-based centroid values provided by \cite{van_der_wel_12}, which is slightly smaller than \cite{Huertas-Company19} cutout size choice ($64{\,\rm pix}\times 64\,{\rm pix}$). Our cutout dimensions correspond to $3\,{\rm arcsec}$ on the side ($\sim 26\,{\rm kpc}$ at $z\sim 2$)  and is at least $\sim 5\times$ the size of $\sim 90\%$ of the galaxies in our sample. We visually randomly inspected our $50{\,\rm pix}\times 50\,{\rm pix}$ cutout images and found that a majority of them encompass the bulk of the galaxy's residual substructure within them. However, it is worth noting that some galaxy's residuals exceed outside our cutout size, which may impact the learning of the DL frameworks. To test for this, we repeated our analysis using a $64\times64$ pix input image size and found no significant differences in the results and conclusions discussed in this chapter. Therefore, we resort to using our initial $50{\,\rm pix}\times 50\,{\rm pix}$ cutout size choice as it is relatively faster to train our model and offers slightly lower dimensionality exploration space for our network.  }}

One possible alternative to our approach (i.e., using object-only residual image) is to use just the {GOI}-only residual image (bottom-center panel of Figure\,\ref{fig:preprocessing}) as inputs to our networks. However, this may impose unforeseen priors on the model learning, where the network focuses more on the shapes of {GOI} segmentation regions. To avoid such biases, we adopt a object-only based approach that is free from the { GOI} segmentation shape information. 

{ Finally, it is a standard practice in the DL-based studies \citep[e.g.,][]{Huertas-Company19} to normalize the input data so that the learning of the deep latent-space vectors are well behaved (i.e., bound and normalized). To conserve the underfit-overfit (i.e., dark-bright structures) appearance of input residual images, we choose to normalize our data to span between $[-1,1]$ using the linear transformation scheme -- {\tt MaxAbs}, where the maximum absolute value of the data sample is set $1$. We tested various available non-linear data transformation schemes\footnote{See \url{https://scikit-learn.org/stable/auto_examples/preprocessing/plot_all_scaling.html\#sphx-glr-auto-examples-preprocessing-plot-all-scaling-py} for more details} (e.g., {\tt PowerTransformer}, {\tt QuintileSclaer}, etc) and found that the choice of data scaling has no impact on the results discussed in this work. }

\subsection{Training-Testing Samples and Data Augmentation}\label{sec:train_test}
\begin{figure}
    \centering
    \includegraphics[width=1\columnwidth]{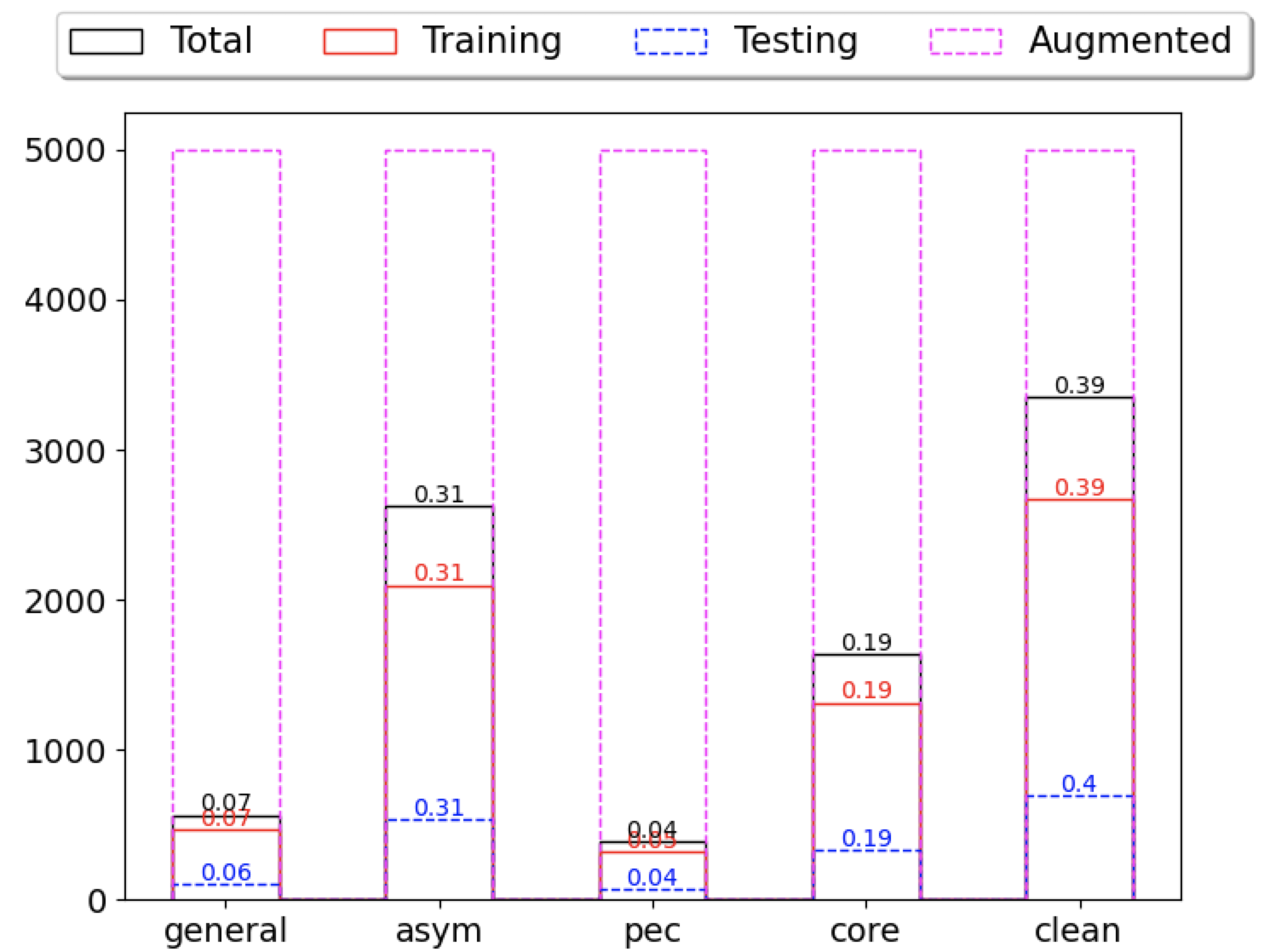}
    \caption[Residual characteristic class-wise distributions of the parent, training, and testing samples]{The histograms of our five residual characteristic classes among our parent sample of $\sim 10,000$ galaxies (black), training sub-sample (red), testing sub-sample (blue), and augmented training data sets (magenta). For each class, we also show its fractional contribution to the data set with respective colored text.}
    \label{fig:train_test_splitting}
\end{figure}

\begin{figure}
    \centering
    \includegraphics[width=1\columnwidth]{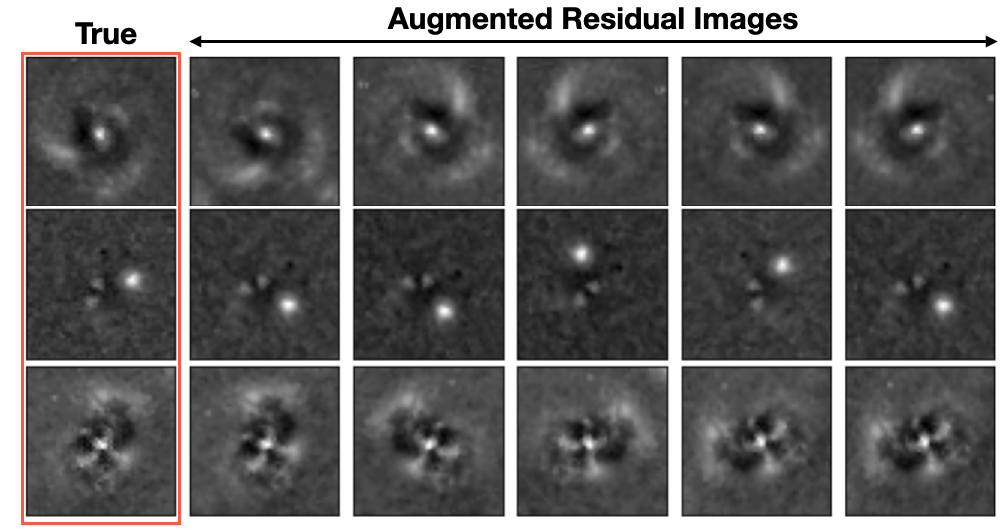}
    \caption[Example visualization of data augmentation process]{Example visualization of our data augmentation step (see \S\,\ref{sec:train_test}). For three example residual images from our sample, we show five random horizontal flip and $45\deg$ rotation steps.}
    \label{fig:data_aug}
\end{figure}

{We sub-divide our main sample information of $10,046$ galaxies and their corresponding object-only images into training and testing sub-samples, while excluding the objects with possible fit-quality issues, using a commonly-employed splitting ratio of $80\%-20\%$, yielding $7948$ training and $1987$ testing galaxies. In Figure\,\ref{fig:train_test_splitting}, we show the class-wise distribution of our total parent sample, subdivided into training and testing samples. We also quote the fractional contribution of each class towards the data sets. From this figure, we illustrate the class-wise proportionate sampling of the training and testing data from the main parent sample. However, { a class-wise imbalance in the parent sample, where certain classes (e.g., Clean, Asymmetric) dominate in numbers}, propagates into the training and testing sample construction. Such an imbalance in the training sample will lead to class-specific biasing of the DL model learning. Furthermore, despite our galaxy sample hosting a diversity of the residual substructures (see Figure\,\ref{fig:residual_visualization}), inclination-induced effects (e.g., similar substructure viewed at different inclinations) can induce a dataset-specific bias and a lack of stochasticity at the deep latent-space level, which may limit their generalized applicability to other data sets. Data augmentation using sample-specific invariant properties (e.g., rotation, brightness/intensity) is a popular choice in the literature to alleviate from such biases \cite[e.g., see][]{Dieleman15,Huertas-Company18}. Conceptually, data augmentation provides a robust low-level learning by inducing a data-level stochasticity,  by generating additional training samples perturbed randomly by changing one or more of their invariant properties.

Motivated by a similar methodology employed in \cite{Huertas-Company19}, we adopt the data augmentation step on our training (object-only) image sample such that each image is augmented with a random horizontal flip and random rotation by $45\deg$. We generate our ``augmented'' training image sample such that each residual characteristic class contains $5000$ images, which are randomly sampled from a large pool of augmented realizations of our parent sample object-only images. In total, our augmented training sample amounts to $25000$ images.  In Figure\,\ref{fig:data_aug}, we show an example visualization of this data augmentation step for example residuals from our training sample. In Figure\,\ref{fig:train_test_splitting}, we also show the uniform class-wise distribution of augmented training sample alongside the lopsided, non-augmented raw training set. We use the augmented training images to train both of our supervised and unsupervised DL networks.

{ It is worth noting that although our augmentation step enables data-level stochasticity, it does not alleviate the underlying human subjectivity of residual characterization, which is especially relevant to our supervised learning case. Recent advancements in the DL literature has proposed various ways to account for human classifier biases and errors. For e.g., Bayesian convolutional neural networks \citep[e.g., ][]{Walmsley20} have the capability to incorporate the biases during learning phase and propagate it through to the output posterior class-probability distributions. It is beyond the scope of this current work to investigate the role of human subjectivity on our analysis results, and we leave this for future work. }  }
 
\subsection{Residual Quantitative Metrics} \label{sec:res_quant}
In addition to the visual-based residual characterization information, we also derive additional { residual quantification metrics: {\it Significant Pixel Flux} ({\it SPF}), commonly-used quantities in the literature  -- {\it Bumpiness} \citep[{\it B};][]{Blakeslee06}, and {\it Residual Flux Fraction}  \citep[{\it RFF};][]{Hoyos12}}  to assess the extracted latent space of our supervised and unsupervised approaches and independently interpret their learning { (see \S\,\ref{sec:results})}. We describe the derivation of these metrics for our parent galaxy sample as the following.

\begin{figure}
    \centering
    \includegraphics[width=1\columnwidth]{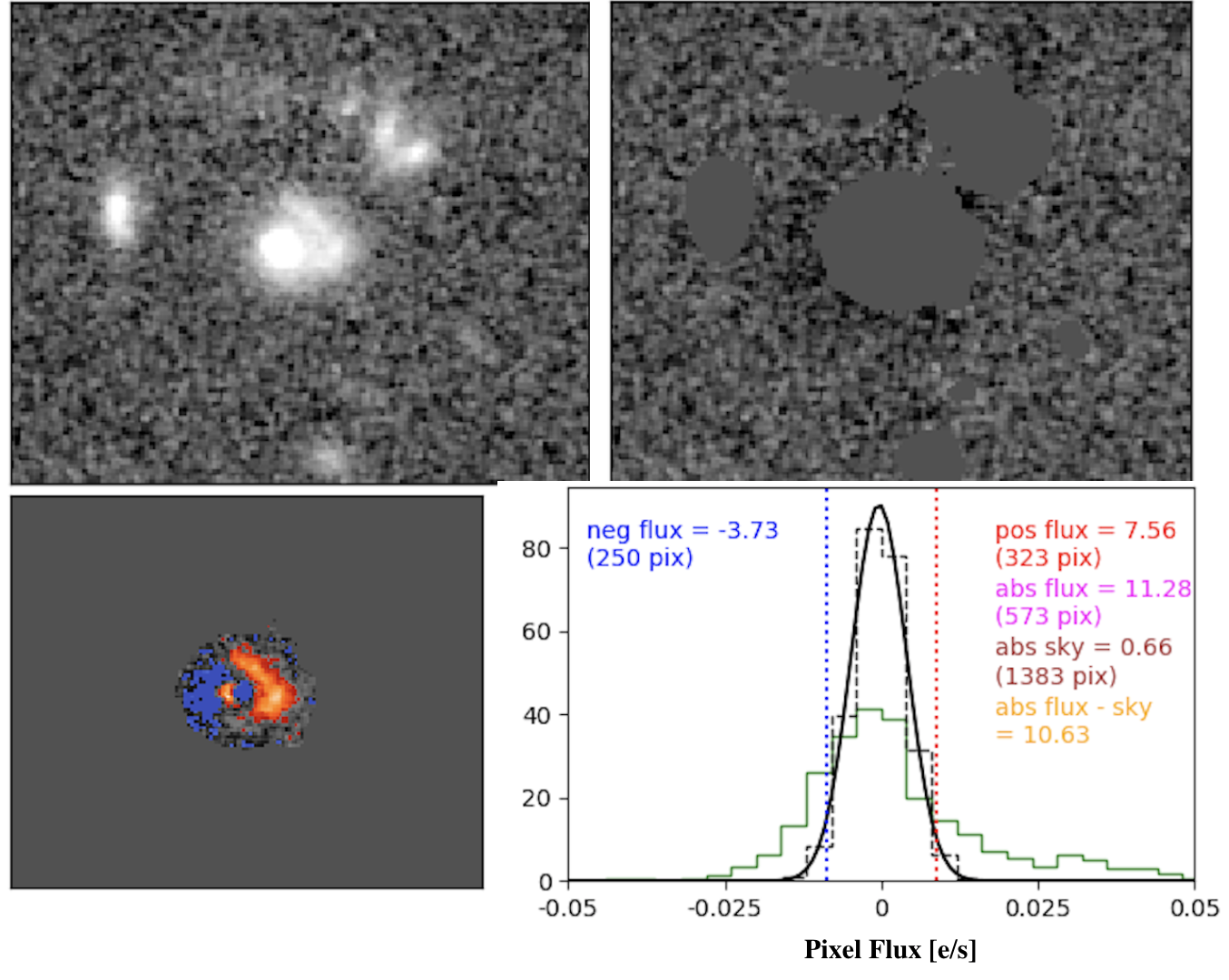}
    \caption[Visual illustration of the significant pixel flux ($SPF$) calculation]{Visual illustration of our Significant Pixel Flux ({\it SPF}) calculation using an example galaxy image (top-left) from our sample. We show the source-masked image (top-right) and the Region of Interest ({\it ROI}; bottom-left) highlighting the Galaxy of Interest ({\it GOI}). We show the histograms of pixel fluxes (bottom-right) in the ROI (green), the sky background distribution (dashed black) fit with a Gaussian (solid black) and their lower and upper $3\sigma$ bounds with blue and red vertical lines, respectively. We highlight those pixels that are above and below the $3\sigma$ bounds in bottom-left panel with their respective colors. For this example, {\it SPF}$=11.28$.}
    \label{fig:spf_calc}
\end{figure}

The cumulative flux of significant pixels above the sky background serves as a straight-forward quantitative indicator for the strength of left-over residual light. As such, for each galaxy in our sample, we quantify the {\it SPF} among the pixels corresponding to the GOI that satisfy $>3\sigma$ significant above the background as:
\begin{equation}
    SPF = \sum_{x,y\in GOI} |~f_{\rm pix>3\sigma} (x,y)~|,
\end{equation}
where $f_{x,y}$ is the flux of the pixel corresponding to pixel coordinate $(x,y)$, and the $GOI$ region is the product of the binary GOI mask (top-right panel in Figure\,\ref{fig:preprocessing}) and the residual data. In Figure\,\ref{fig:spf_calc}, we show an example visual illustration of the {\it SPF} calculation. For each galaxy, we identify sources in the original image using a python-based implementation {\tt SEP} \citep{barbary16} of a standard source-extraction software {\tt SExtractor} \citep{bertin96}.  We generate a binary sky mask such that all the source segmentation regions are set to zero and the background sky regions are set to unity (top-right panel in Figure\,\ref{fig:spf_calc}). We then generate a binary {\it GOI} mask as described in \S\,\ref{sec:train_test} and produce a GOI residual region by multiplying the {\it GOI} mask with the residual data (bottom-left panel in Figure\,\ref{fig:spf_calc}). We extract the distribution of background sky pixels from the background sky image (sky mask $\times$ residual image), and fit it with a Gaussian to derive its upper and lower $3\sigma$ bounds. We select the pixels in GOI residual region that have $f (x,y)>3\sigma$ and $f(x,y)<3\sigma$, and derive its {\it SPF} value as the cumulative absolute flux sum of these pixels. For the example shown in Figure\,\ref{fig:spf_calc}, the total positive flux of $>3\sigma$ pixels is $7.56\,{\rm e/s}$ (native, exposure time normalized image units) and the total negative flux is $-3.73\,{\rm e/s}$, which corresponds to a $SPF = 11.28\,{\rm e/s}$.

We also quantify the {\it Bumpiness} \citep{Blakeslee06} parameter, which conceptually quantifies the strength of second-order light-profile moments and deviations of galaxy's structure from a smooth parametric model. It is defined as the ratio of the residual rms to the best-fit s\'ersic mean model:
\begin{equation}
    B = \frac{\sqrt{(1/N) ~\sum_{(x,y)\in GOI}~[R_{s}^{2}(x,y) - \sigma_{\rm sky}^{2}]}}{(1/N) ~\sum_{(x,y)\in GOI}~[S(x,y)]},
\end{equation}
where $R_{s}$ is the smoothed residual image after convolution with a  2D BoxCar filter of size $3\times3\,{\rm pix}$, $\sigma_{\rm sky}^{2}$ is the variance of background sky computed using the best-fitted Gaussian to the background-sky distribution in the above $SPF$ calculation steps, $S$ is the best-fit s\'ersic model generated by \cite{van_der_wel_12}, and $N$ is the number of pixels corresponding to the GOI segmentation region (i.e., white regions in GOI-mask panel of Figure\,\ref{fig:preprocessing}).

Finally, we also compute the {\it Residual Flux Fraction}  \cite[{\it RFF};][]{Hoyos12}, which conceptually quantifies the additional residual light that cannot be accounted by background noise fluctuations. It is defined as:
\begin{equation}
    RFF = \frac{\sum_{x,y\in GOI} R(x,y) - 0.8 \times \sum_{x,y\in GOI} \sigma_{\rm sky}}{\sum_{x,y\in GOI} S(x,y)},
\end{equation}
where $R$ is the residual image, $0.8\times \sum \sigma_{\rm sky}$ is the expectation of residual light from Gaussian noise fluctuations, and $S$ is the best-fit light profile model. In \S\,\ref{sec:results}), we assess the deep latent space features of our trained DL models in conjunction with $SPF$, $B$, and $RFF$ metrics to interpret their learning.

\begin{figure*}
    \centering
    \includegraphics[width=2\columnwidth]{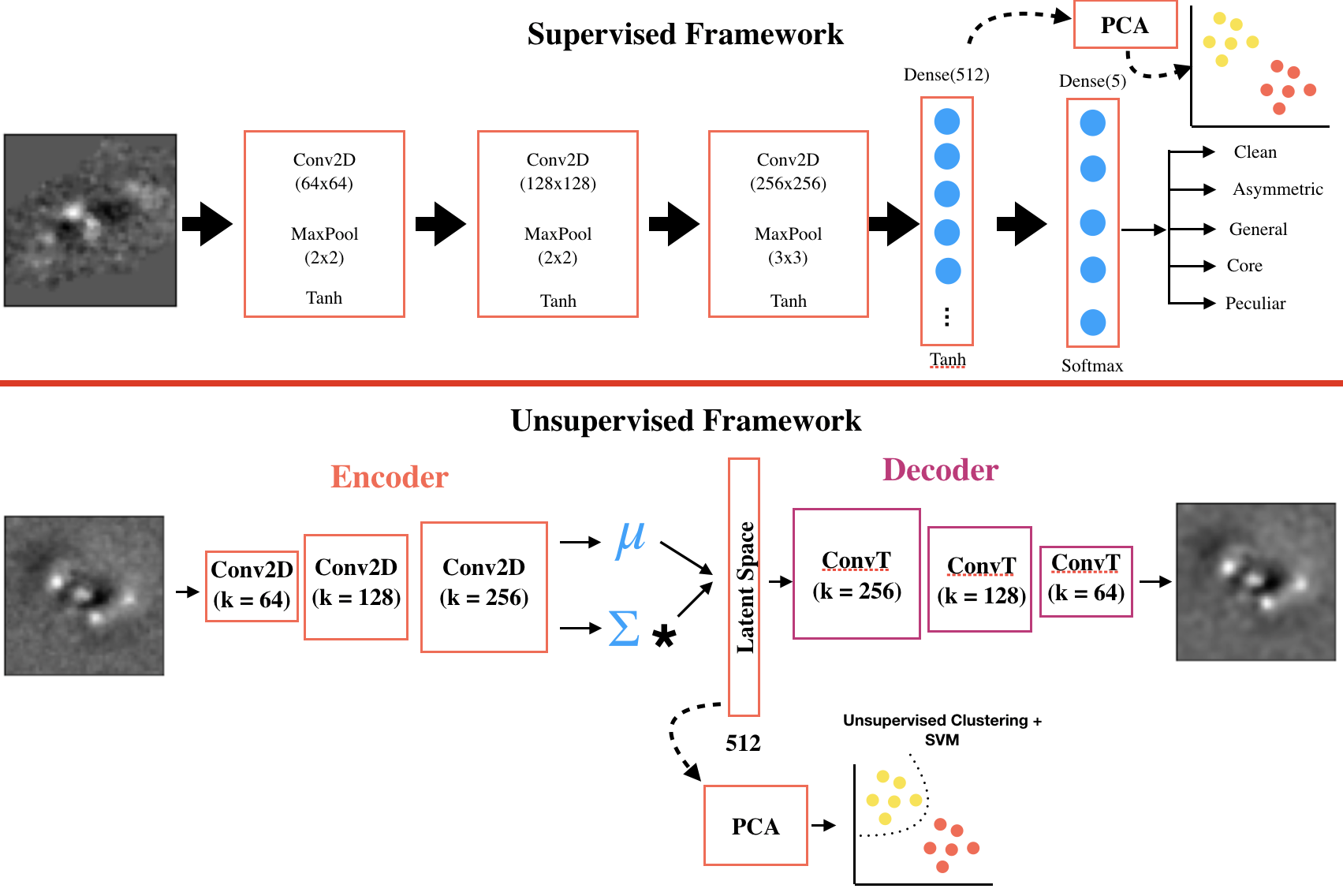}
    \caption[Overview of the Supervised CNN and Unsupervised Varitational Autoencoder Architectures]{Overview of our supervised CNN (top) and unsupervised Convolutional Variational Autoencoder (bottom) frameworks used to analyze the galactic residual substructures (see \S\,\ref{sec:cnn_model} \& \S\,\ref{sec:cvae_model}). For both networks, we also visually illustrate our analysis of the latent space features using Principle Component Analysis and unsupervised clustering (see \S\,\ref{sec:results}).}
    \label{fig:architecture}
\end{figure*}

\section{Proposed Deep Learning Models}\label{sec:models}
{ To better automate the characterization of residual galactic substructures and improve the identification of plausible merging signature hosting galaxies, we develop two frameworks -- i) a supervised Convolutional Neural Network (CNN) model; ii) an unsupervised  Convolutional variational Autoencoder (CvAE) model -- to learn the latent space features describing the variety of residual substructures hosted by our sample of $\sim 10,000$ massive galaxies from the {\it HST} CANDELS survey.  In this section, we describe the architecture of our CNN and CvAE frameworks and discuss their training and optimization strategy.}

\subsection{Supervised Framework}\label{sec:cnn_model}
The architecture of our supervised framework is inspired from the CNN-based classifier developed by \cite{Huertas-Company15}.  In Figure\,\ref{fig:architecture}, we show the visual overview of our supervised CNN-based framework. Briefly, our network takes the ``object-only'' residual image stamps as single channel inputs (i.e., one object-only image per sample), followed by three sets of Convolutional and MaxPooling layers for feature extraction, and two Dense Fully-Connected layers. Given that our input data is scaled and spans [-1,1], we choose a ``{\it Tanh}'' activation function for all the convolutional layers, whose resultant activations also span [-1,1]. For our deep latent space feature layer, we choose to extract $512$ vectors, motivated by a separate dimensionality reduction exercise, where we apply a direct PCA decomposition of the object-only training images and find that $512$ eigen vectors hold $\gtrsim 90\%$ of the explained sample variance (i.e., $>90\%$ of information is captured by reduced, $512$ dimensions). We choose {\it Tanh} activation for this Dense layer and a softmax activation for the final classification layer (of size $5$) such that the output is probabilistic prediction per target class (here visual residual class labels) that sums up to unity. Following the standard practices to avoid over-fitting scenarios, we introduce a 50\% dropout layer after each Convolutional$+$MaxPooling block, which conceptually omits (randomly) half of the layer features from propagating further. We experimented with different dropout percentage values and found that using lower values (e.g., $\sim 5-10\%$) resulted in worsened model performance with reduced testing set accuracy due to overfitting. Furthermore, we also introduce a Gaussian Noise layer with a standard deviation $\sigma =0.02$ at the entrance of the network to facilitate learning of latent space that is less susceptible to image noise patterns, and ensures that the network never sees the exact same image during model training. Using a higher value for the entrance random noise layer (e.g., $\sigma \sim 0.1$) induces too much stochasticity into the network and leads to an oscillating model convergence (i.e., cyclic learning-unlearning patterns). 

To optimize our supervised CNN model during the training, we use {\it categorical cross-entropy} loss function with {\it Adam} optimizer \citep{kingma17} with starting learning rate of $10^{-4}$. We train our network with the augmented training sample of $25000$ images for a maximum of $100$ epochs in batch sizes for weight updation of $32$ images, and decay the learning rate by $2\times$ for every $1/3^{\rm rd}$ of the total steps to aid the model convergence. Our model achieved an accuracy of $\sim 95\%$ after $75$ training epochs and it yielded a testing accuracy of $\sim 75\%$. We note that the model achieves an accuracy of $\sim 91\%$ on the Clean class, but the overall (lower) testing accuracy is due to considerable cross-class confusion between the Peculiar, General, and the Asymmetric classes. We acknowledge that this stage needs further improvement and will one of the main items of future work. In the later section, we assess the latent space features learned by our CNN model.

\subsection{Unsupervised Framework}\label{sec:cvae_model}
There are several approaches to learning the deep latent space in an unsupervised fashion, amongst which Autoencoders, specifically Convolutional Autoencoders \citep[CAE;][]{Masci11} are a popular option. Briefly, they comprise of an Encoder-Decoder framework, where the objective is to learn the dimensionality-reduced latent space vectors via iterative adjusting of layer weights such that the output reconstruction (by a Decoder network) matches closely to the input image. Motivated by this ideology, we implement an unsupervised Convolutional Autoencoder based framework to learn the deep latent space that characterizes different galactic residual substructures. Specifically, we adopt a Convolutional Variational Autoencoder (CvAE), which is an improvement over the simple CAE, where the network learns a distribution of latent space vectors described by a unit Gaussian distribution instead of simple latent space point vector, which makes the learned latent space less sensitive to unknown nuisance in the input images. In practice, we noticed that the output reconstruction from the CvAE is qualitatively (i.e., by visually by eye) closer to the input images than CAE counterparts (which are often more ``blurry'' in appearance). 

We model our CvAE framework based on the previously mentioned supervised CNN architecture (see \S\,\ref{sec:cnn_model}). In Figure\,\ref{fig:architecture} (bottom panel), we show our CvAE architecture,  where our Encoder network involves three 2D Convolutional layers, whose end features are flattened and passed to a fully-connected, latent space feature layer of size $512$ (following our aforementioned PCA-based suggestions). We then connect this latent feature layer to two more fully-connected layers (each of size $512$) to describe the mean ($\mu_{l}$) and variance ($\sigma^{2}_{l}$) of the latent space features ($l$), which are then passed to a custom unit Gaussian Sampling layer. The output of this layer (and the encoder network) is a latent space feature vector of size $512$, where each feature is sampled from the Gaussian($\mu_{l}$,$\sigma^{2}_{l}$). 

Our decoder network starts with an input layer of size $512$ (i.e., matching the Encoder output size), followed by three 2D transpose convolutional layers that mirror the sizes of our encoder networks' convolutional layers. We use {\it ReLu} activation function for all the layers in the Encoder-Decoder framework, except for the output layer of the decoder network where we use the {\it Tanh} activation. Finally, similar to our CNN framework, we also introduce a Gaussian Noise layer ($\sigma = 0.02$) at the start of the encoder framework to enable latent-space features that are less susceptible to image noise patters. It is worth noting that we do not use feature dropout in between the convolutional layers (unlike in our CNN framework), as in doing so, the output reconstructions qualitatively deviated considerably from their input counterparts, where dropping features resulted in blurrier reconstructions.

\begin{figure}
    \centering
    \includegraphics[width=\columnwidth]{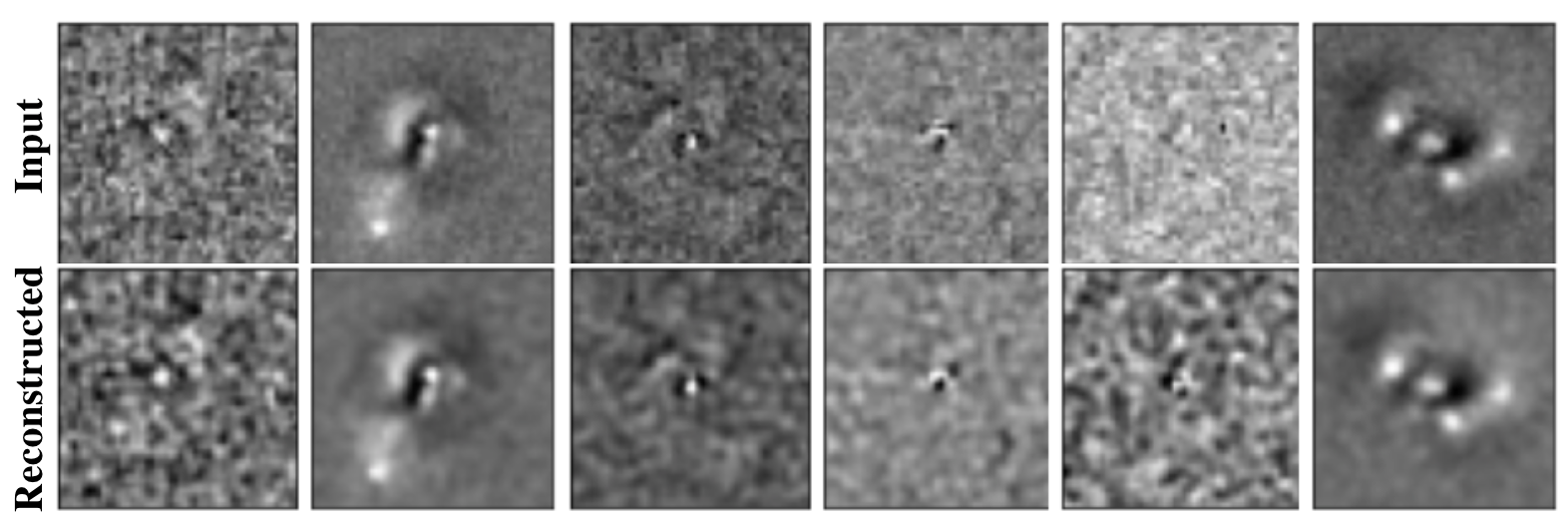}
    \caption[Comparative visualization of the input vs. reconstructed residual images by the unsupervised variation autoencoder framework]{Example visualization of our unsupervised Convolutional Variational Autoencoder network's input (top) and reconstruction (bottom) images.}
    \label{fig:vae_recon}
\end{figure}

To optimize our network during the training, we jointly minimize the input image {\it reconstruction loss} and the {\it Kullback-Leibler} ({\it KL}) divergence loss of the latent space feature distribution. Conceptually, our optimization process corresponds to requiring the reconstructed outputs to be closer to the input images while simultaneously ensuring that their underlying latent space features (describing the input-output images) are also closely matched. Our {\it total loss} function, which is the summation of the image reconstruction loss and the {\it KL}-divergence loss is as follows:

\begin{equation}
    \mathcal{L}_{\rm total} = \mathcal{L}_{\rm recon} + \mathcal{L}_{\rm KL},
\end{equation}
where we choose $\mathcal{L}_{\rm recon}$ as the mean-squared error ({\it MSE}) between the input ($I$) and the reconstructed image ($\hat{I}$) as:
\begin{equation}
    \mathcal{L}_{\rm recon} = \frac{1}{N} \sum_{i,j} (I - \hat{I})^{2}.
\end{equation}
The {\it KL} divergence loss of the latent space vectors ($l$) with  mean ($\mu_{l}$) and variance ($\sigma_{l}^{2}$) is given by:
\begin{equation}
   -0.5 \times [1 + \log_{10}(\sigma_{l}^{2}) - \mu^{2}_{l} - \exp\{\log_{10}(\sigma_{l}^{2})\}]~.
\end{equation}
To optimize our CvAE network during the training, we adopt the same strategy as in our supervised approach (see \S\,\ref{sec:cnn_model}) with the {\it Adam} optimizer and an initial learning rate of $10^{-4}$ with a decay factor of two every $1/3^{\rm rd}$ of the steps over a maximum of $100$ training epochs. In Figure\,\ref{fig:vae_recon}, we show the input-reconstructed image pairs of example galaxy residuals from the testing data. We notice that the overall visual structure of reconstructed residuals is preserved, albeit with a globally ``smoothed'' appearance.

\section{Discussion: Assessment of Supervised and Unsupervised Latent Space using PCA and Unsupervised Clustering}\label{sec:results}
Thus far, we have developed two DL frameworks, a supervised CNN and an unsupervised CvAE network, and trained them on our augmented training image sample of residual images to learn the latent space representations of different residual substructure characteristics spanning in our data set. In this section, we assess the supervised and unsupervised latent space features with Principle Component Analysis \citep[PCA;][]{Jolliffe86,Calleja04} along with supplemental, independently derived quantitative metrics ({\it SPF}, {\it B}, and {\it RFF}). We also explore unsupervised clustering algorithms and Support-Vector Machines \citep[SVM; ][]{Hearst98,Huertas-Company11} to identify classification boundaries in the PCA space that can distinguish different residual structures.

\begin{figure*}
    \centering
    \includegraphics[width=2\columnwidth]{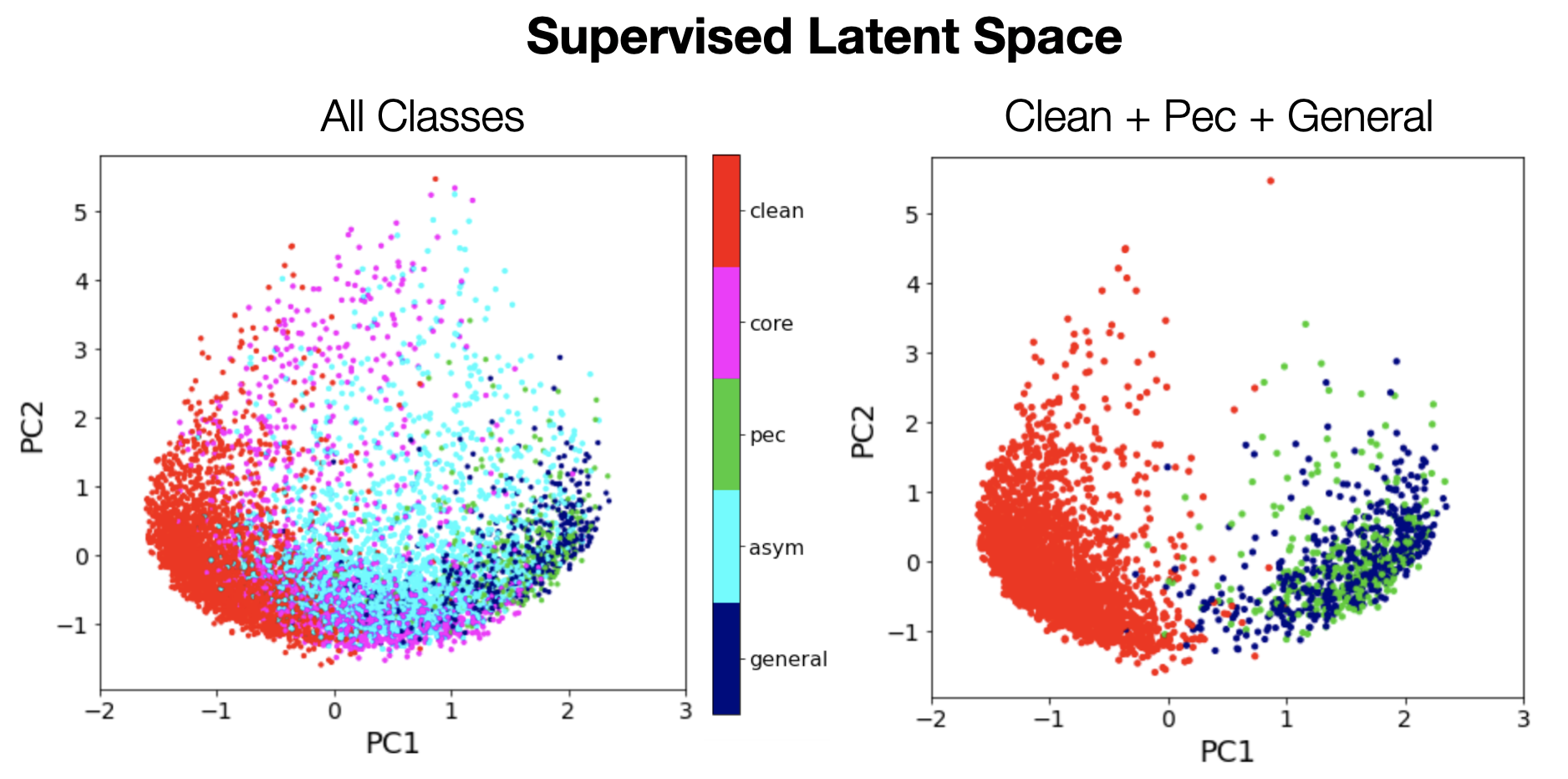}
    \caption[Visualization of the supervised CNN latent space in PCA embedding]{Visualization of principle component axes of the supervised latent space color coded with five residual class labels (left panel). In the right panel, we show the PC1 vs. PC2 only for the Clean, Peculiar, and General Classes. }
    \label{fig:supervised_result}
\end{figure*}

\begin{figure*}
    \centering
    \includegraphics[width=2\columnwidth]{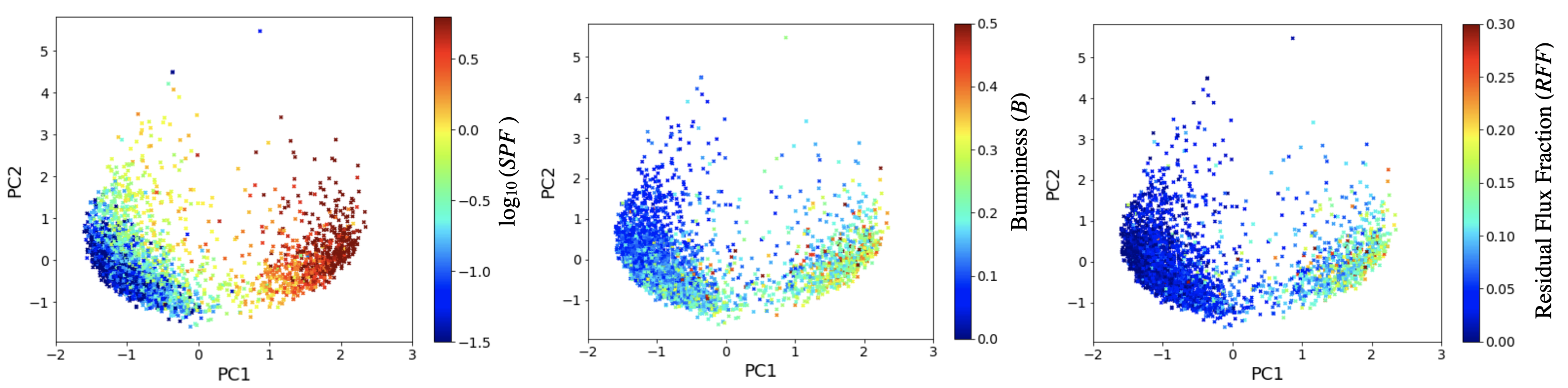}
    \caption[Visualization of the supervised CNN latent space in PCA embedding with supplemental quantitative metrics]{The supervised CNN based latent space visualized in the PCA eigen axes for a subset of classes (Clean, Peculiar, and General; shown in Figure\,\ref{fig:supervised_result}) and color coded by different residual quantitative metrics -- $\log_{10} (SPF)$ (left panel), Bumpiness ($B$; middle panel), and Residual Flux Fraction ($RFF$; right panel).  }
    \label{fig:supervised_pca_SPF_B_RFF}
\end{figure*}

\subsection{PCA Analysis of Supervised CNN Latent Space}\label{sec:supervised_latent_space}
Using our trained CNN network, we extract the latent space features representing the {\it unaugmented} training image set by running them through the network (with frozen, trained weights) and extracting information from the fully-connected layer penultimate to the classification output (see Figure\,\ref{fig:architecture}). We then run PCA on these latent features using the python-based implementation {\tt scikit-learn} \citep{scikit-learn} and extract three top-most information bearing eigen components capturing $\gtrsim 90\%$ of the explained variance. In Figure\,\ref{fig:supervised_result} (left panel), we show the training sample projected in the first two principle component axes (PC1 vs. PC2) and { color code} their corresponding visual residual class labels. We notice that the Clean vs. Peculiar and General classes appear distinctly separate (by eye) in the PCA space, where the bulk of Clean class objects are clustered alongside the ``left-arm'' and the Peculiar and General classes are along the ``right-arm'' in the apparent ``U'' shape of the PC1 vs. PC2 distribution. This distinction can be clearly seen in the right panel of Figure\,\ref{fig:supervised_result}, where we show the PC1 vs. PC2 distribution for peculiar and general classes (i.e., galaxies hosting strong residuals) as these objects are of particular scientific interest, in the context of merging induced tidal features. On the other hand, we notice that the core and asymmetric classes lie in the ``saddle'' region of the PC1 vs. PC2 distribution ($PC1\in\, \sim [0,1]$), and are considerably overlapping with the peculiar and general classes (at $PC1 \gtrsim 1$). { Our results based on Figure\,\ref{fig:supervised_result} conceptually illustrate that the CNN learns a higher-order measure of residual strength (especially along PC1) and uses it as a primary leverage to characterize residual substructures into strong, intermediate, and clean groupings. This is intuitive as the CNN's learning mimics that of a human's interpretation of the residual strengths when characterizing them. }

\begin{figure*}
    \centering
    \includegraphics[width=2\columnwidth]{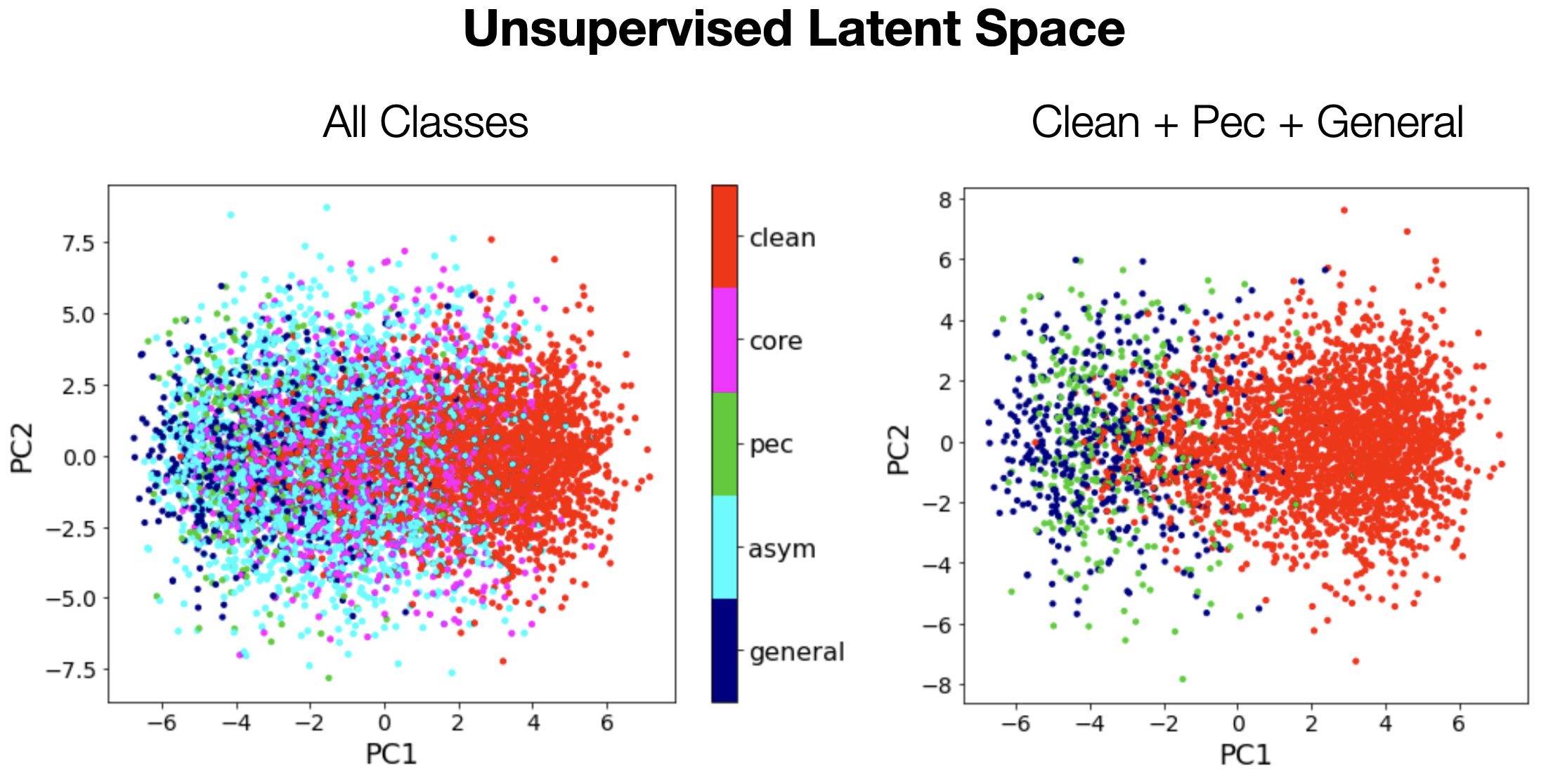}
    \caption[Visualization of unsupervised variational autoencoder latent space in PCA embedding]{Visualization of the unsupervised CvAE latent space in the PCA eigen axes (PC1 vs. PC2), where each data point is color coded by its visual residual class label  (left panel). In the right panel, we only show the Clean, Peculiar, and General classes.}
    \label{fig:unsupervised_result}
\end{figure*}

\begin{figure*}
    \centering
    \includegraphics[width=2\columnwidth]{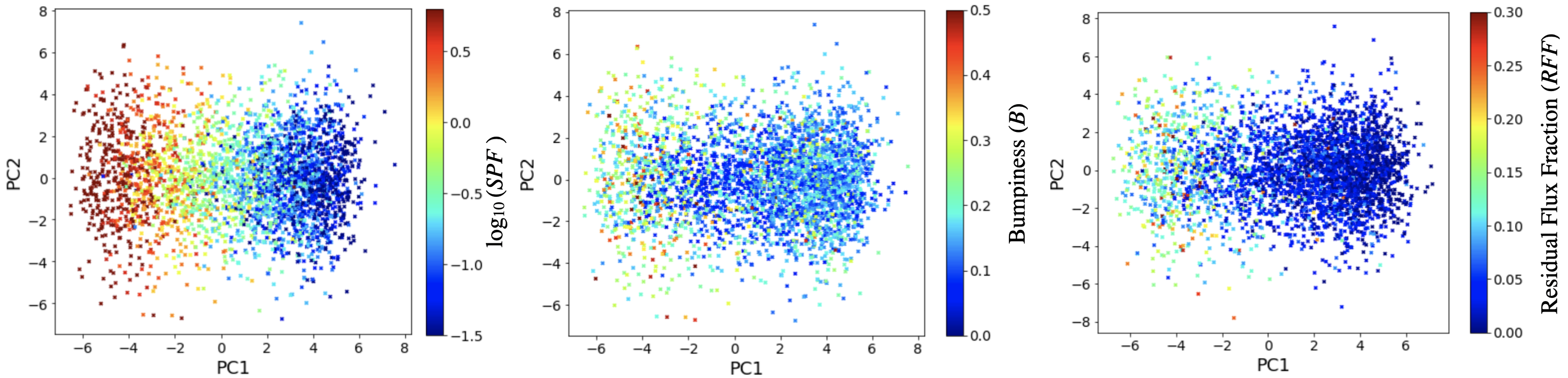}
    \caption[Visualization of unsupervised variational autoencoder latent space in PCA embedding with supplemental quantitative metrics]{Visualization of the unsupervised CvAE latent space in PCA eigen axes along with residual quantitative metrics $SPF$, $B$, and $RFF$ following a similar layout as in Figure\,\ref{fig:supervised_pca_SPF_B_RFF}.}
    \label{fig:unsupervised_pca_SPF_B_RFF}
\end{figure*}

{ To test our above interpretation that the PC1 vs. PC2 may be correlated with a measure of quantitative residual strength, we assess the learned latent space with independently quantified measures of residual strength -- $SPF$, $B$, and $RFF$.} In Figure\,\ref{fig:supervised_pca_SPF_B_RFF}, we show the PC1 vs. PC2 distribution for the Clean, Peculiar, and General residual classes along with their quantitative metrics $SPF$, $B$, and $RFF$. We find that all the three quantities correlate { distinctly} with the with the PC1 and weakly with the PC2 axis. More notably, we find a strong bimodal correlation of PC1 and PC2 with the $SPF$, where the objects with $PC\gtrsim 0.5$ have higher $SPF$ with $\sim 10-100\times$ than the objects at $PC1\lesssim 0$ ($\log_{10}(SPF) \sim -0.5\,{\rm to}\,-1.5$). We also note a relatively weaker, but noticeable correlation with the Bumpiness and Residual Flux Fractions, where the $PC\gtrsim 0.5$ have higher $B$ and $RFF$ values than the objects spanning $PC\lesssim 0$. { Our observations based on  Figure\,\ref{fig:supervised_pca_SPF_B_RFF} conceptually suggest that the latent space learned by our CNN network physically corresponds to a combination of different residual strength metrics.}

{ It is worth noting that the training sample distribution along the third principle component (PC3) is nearly identical to the PC2 distribution. As a result, our analysis of the above trends in PC1 vs. PC3 space are very similar to the PC1 vs. PC2 based conclusions.  Therefore, we just focus on the PC1 vs. PC2 space to make our main points.}
\begin{figure*}
    \centering
    \includegraphics[width=2\columnwidth]{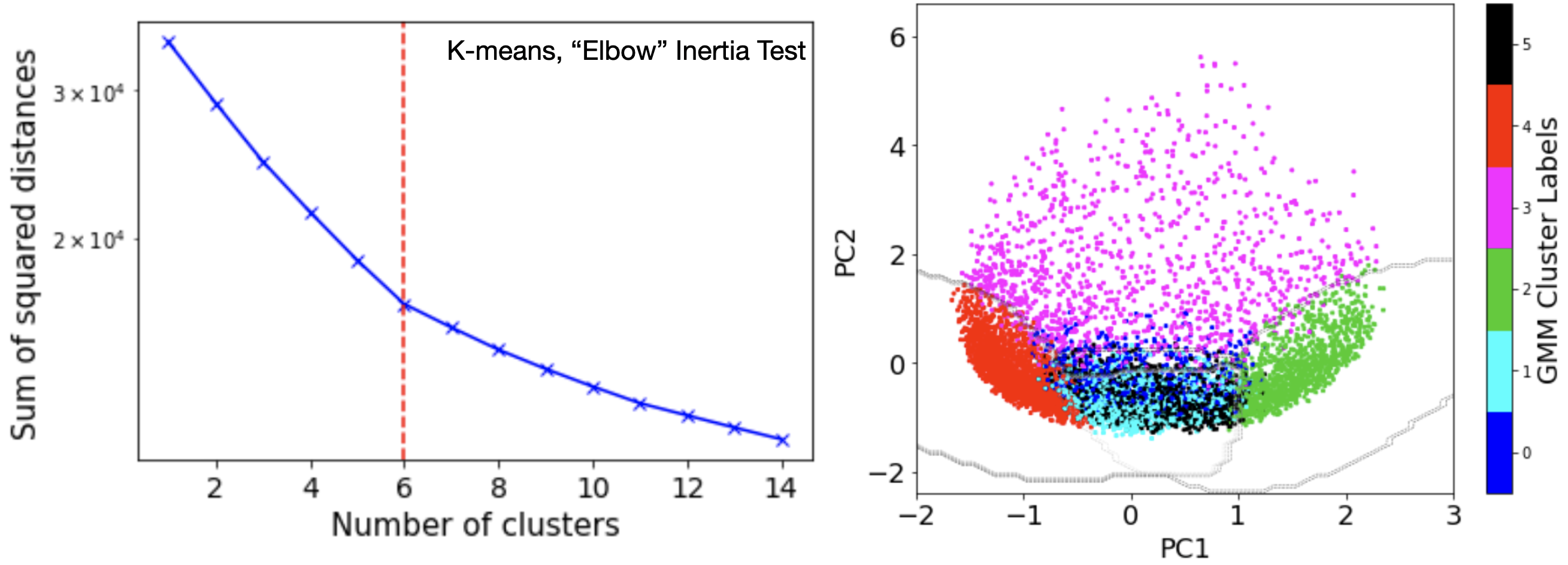}
    \caption[Gaussian Mixture Model (GMM) based unsupervised clustering of the CNN latent space ]{{\it Left panel}: The sum of squared distances between the data points using $k$-means clustering technique as a function of number of permissible clusters, indicating a transition from a sharp to gradual fall off at an optimal number of $6$ clusters. {\it Right panel}: We show unsupervised Gaussian Mixture Modeling based clustering of the PCA embedding of the CNN latent features, where the GMM identified clusters are labelled with unique color coding (see color bar).}
    \label{fig:supervise_pca_gmm}
\end{figure*}

\begin{figure}
    \centering
    \includegraphics[width=1\columnwidth]{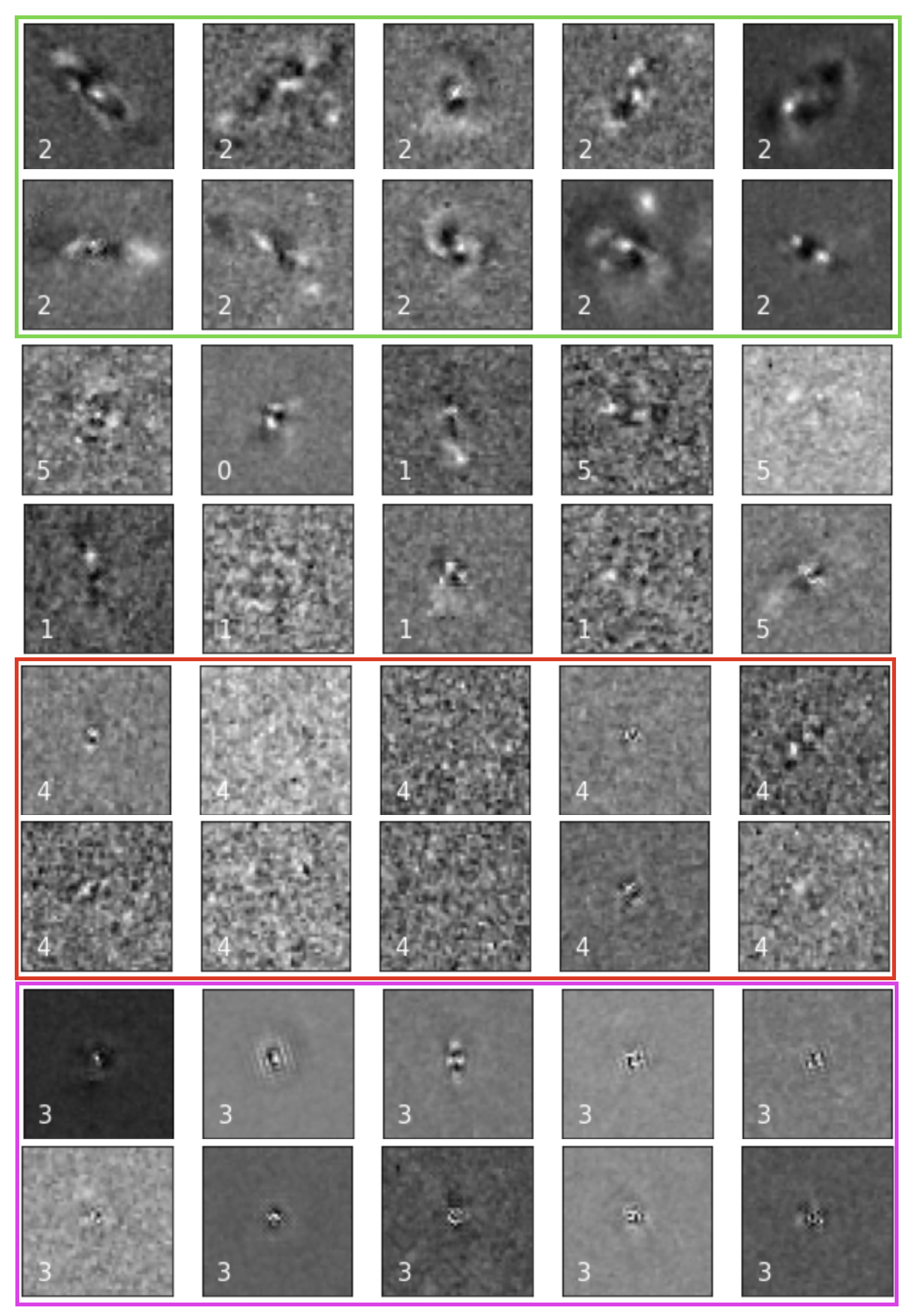}
    \caption[Visualization of residuals in the GMM clustered supervised CNN latent space ]{Visualization of the residual images  sampled randomly from the GMM-based clusters identified in the supervised CNN-based PCA space (in Figure\,\ref{fig:supervise_pca_gmm}). We outline the row pairs with borders corresponding to the color coding used in  Figure\,\ref{fig:supervise_pca_gmm}, except for the second and third rows, which are random samplings from multiple clusters (GMM labels $= 0,1,5$). On each residual image, we indicate the GMM cluster label.}
    \label{fig:supervise_gmm_res}
\end{figure}

\subsection{PCA Analysis of Unsupervised CvAE Latent Space}\label{sec:unsupervised_latent_space}
Using our encoder frontend of the trained unsupervised CvAE network, we extract the latent space features corresponding to our unaugmented training image sample. Analogous to our supervised case, we then perform PCA analysis on these latent vectors and analyze trends in the first two principle components PC1 and PC2. In Figure\,\ref{fig:unsupervised_result} (left panel), we show the PC1 vs. PC2 distributions for all five visual residual classes, and only focus on the Clean, Peculiar, and General classes in its right panel. Immediately, we notice that the overall structure of the PC1 vs. PC2 distribution is less distinctive and more contiguous than our supervised case, where the Clean, Core, and Asymmetric classes follow a continuum with no distinguishing correlation across PC1 and PC2 axes. { These results conceptually suggest that although our unsupervised CvAE framework learns a residual strength equivalent measure, it lacks the additional discriminatory information from the visual-based residual characterizations. Nevertheless, it is worth noting that the distribution of General and Peculiar classes along the PC1 axis fall towards the lower tail-end ($PC\lesssim -2$) of the gaussian-like distribution of the Clean class objects.}

Similar to our supervised latent space exercise discussed in \S\,\ref{sec:supervised_latent_space}, we also assess the unsupervised latent features in PC1 vs. PC2  with the residual quantitative metrics ($SPF$, $B$, and $RFF$). In Figure\,\ref{fig:unsupervised_pca_SPF_B_RFF}, we show the PC1 vs. PC2 for our Clean, Peculiar, and General Classes and color code each data point based on their $SPF$, $B$, and $RFF$ values. We find a bimodality in the $SPF$ values along the PC1 axis, where $PC1\gtrsim 2$ ($PC1\lesssim -2$) have smaller (larger) $SPF$ values, where there is continuum of objects with intermediate $SPF$ value during $-2\lesssim PC1 \lesssim 2$. Similarly, we notice a weak correlation between the $B$ and $RFF$ quantities and $PC1$, where smaller $PC1$ values correspond to higher $B$ and $RFF$ measures. It is worth mentioning that there is no distinct correlation between the three metrics and the $PC2$ axis. Our exercise illustrates that the unsupervised CvAE latent space is only somewhat informative (relative to our supervised case) in terms of distinguishing different residual characteristics, where only one informative eigen axis ($PC1$) correlates with (independent) physical measures of residual strength.

\subsection{Unsupervised Clustering of the CNN and CvAE Latent Features in PCA Space}
To assess the discriminatory power of our supervised and unsupervised latent space features { to naturally distinguish between different residual characteristics}, we investigate unsupervised clustering in their corresponding PCA space along with Support Vector Classifiers \citep[SVC;][]{Boser92}. There are various popular unsupervised clustering techniques to identify data-driven natural clusters in N-dimensional parameter space such as {\it k}-means clustering \citep{lloyd57,macqueen67}, Spectral Clustering \citep{alpert99}, Density based spatial clustering \citep[DBscan; ][]{Ester96}, and Gaussian Mixture Modeling \citep[GMM;][]{mclachlan88}. For our exercise, we choose the posterior probability based cluster assignment approach -- GMM, informed by a preliminary $k$-means clustering assessment of the data distribution in PC1 vs. PC2 space. 

\begin{figure*}
    \centering
    \includegraphics[width=2\columnwidth]{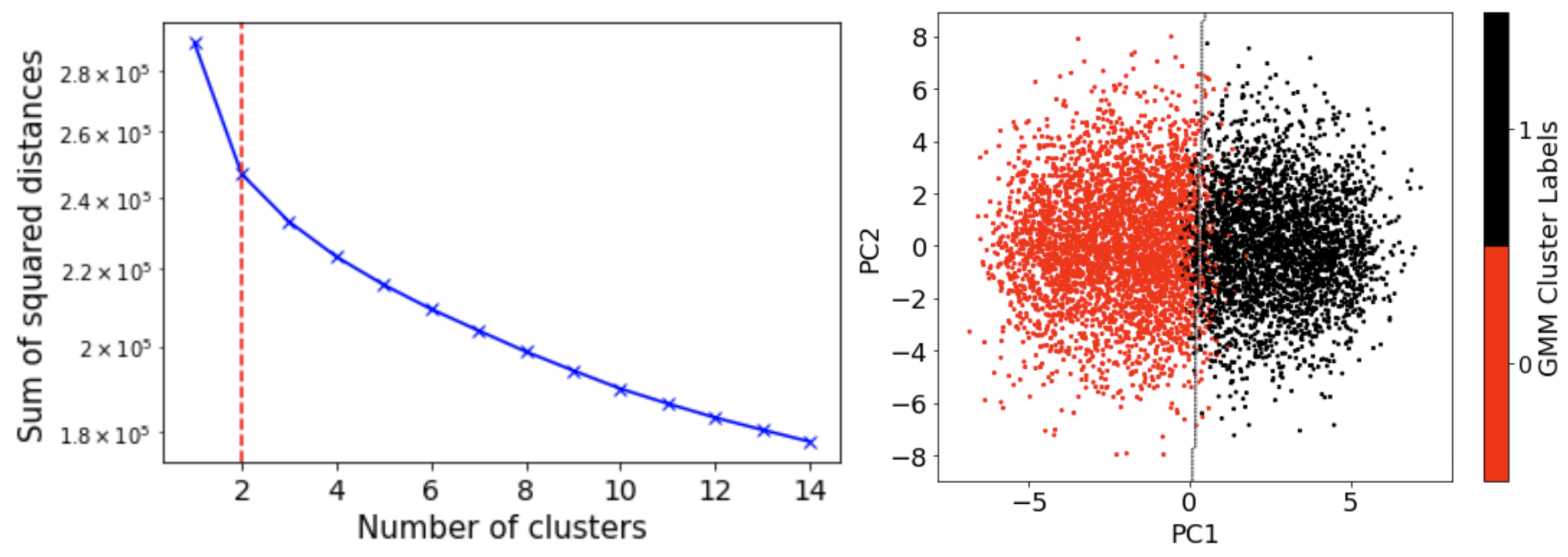}
    \caption[Gaussian Mixture Model (GMM) based unsupervised clustering of the CvAE latent space ]{Same as Figure\,\ref{fig:supervise_pca_gmm}, but for our unsupervised CvAE latent features in PCA space.}
    \label{fig:unsupervise_pca_gmm}
\end{figure*}

First, we repeatedly run $k$-means clustering on the principle components of the supervised CNN latent features with different number of permissible clusters ($k \in [1,15]$) and compute the ``inertia'' as the sum of squared distances of the data points to their nearest cluster centers. In Figure\,\ref{fig:supervise_pca_gmm} (left panel), we show the $k$-means inertia for different number of clusters, which falls sharply up to an optimal value of $k=6$ clusters and shifts to a more gradual fall off. This test to find data-driven optimal number of clusters is dubbed the ``Elbow'' method. Using this information, we run the unsupervised GMM clustering on our CNN latent features in PC1 vs. PC2 space and identify $6$ clusters.

In Figure\,\ref{fig:supervise_pca_gmm} (right panel), we show the PC1 vs. PC2 space with data points color coded based on their GMM identified cluster labels ($n_{\rm gmm,cnn}$). We notice that the clusters $n_{\rm gmm,cnn} = 2,4$ have been naturally identified, which quantitatively correspond to residuals with high and low $SPF$ values, respectively. We also find that the clusters $n_{\rm gmm,cnn} = 0,1,5$ span the intermediate region between the  $n_{\rm gmm,cnn} = 2,4$ clusters, which also quantitatively correlate to intermediate $SPF$ values. Finally, our method identified the $n_{\rm gmm,cnn} = 3$ cluster data as ``outliers'' to the remaining clusters. To visually assess the GMM-based clusters, in Figure\,\ref{fig:supervise_gmm_res}, we show the residual images of the randomly selected GMM-based cluster members. Interestingly, we notice that $n_{\rm gmm,cnn}=2$ indeed hosts galaxies with strong and often interesting residual characteristics, and the residuals of the $n_{\rm gmm,cnn}=4$ objects are nearly ``clean'' in appearance. On the other hand, we notice that the residuals of objects spanning the clusters $n_{\rm gmm,cnn} = 0,1,5$ host weaker (compared to $n_{\rm gmm,cnn}=2$), diffuse asymmetric signatures. Finally, we find that a majority of the $n_{\rm gmm,cnn}=3$ residuals host distinct (yet visually-similar) central residual patterns, with no obvious signatures outside the central region. Our exercise suggests that the supervised CNN model learns latent space which can be segregated into qualitatively and quantitatively similar residual characteristics. To automatically identify quantitative boundaries amongst the identified GMM clusters, we use the SVC with a radial-basis function kernel ($\gamma =0.7$ and $C=1$). In Figure\,\ref{fig:supervise_pca_gmm}, we show the SVM-based classification boundaries of the GMM clusters. Our GMM-based cluster identities and SVM-based classification boundaries can be further used to refine/validate our visual residual class characteristics used for training purposes.

\begin{figure}
    \centering
    \includegraphics[width=1\columnwidth]{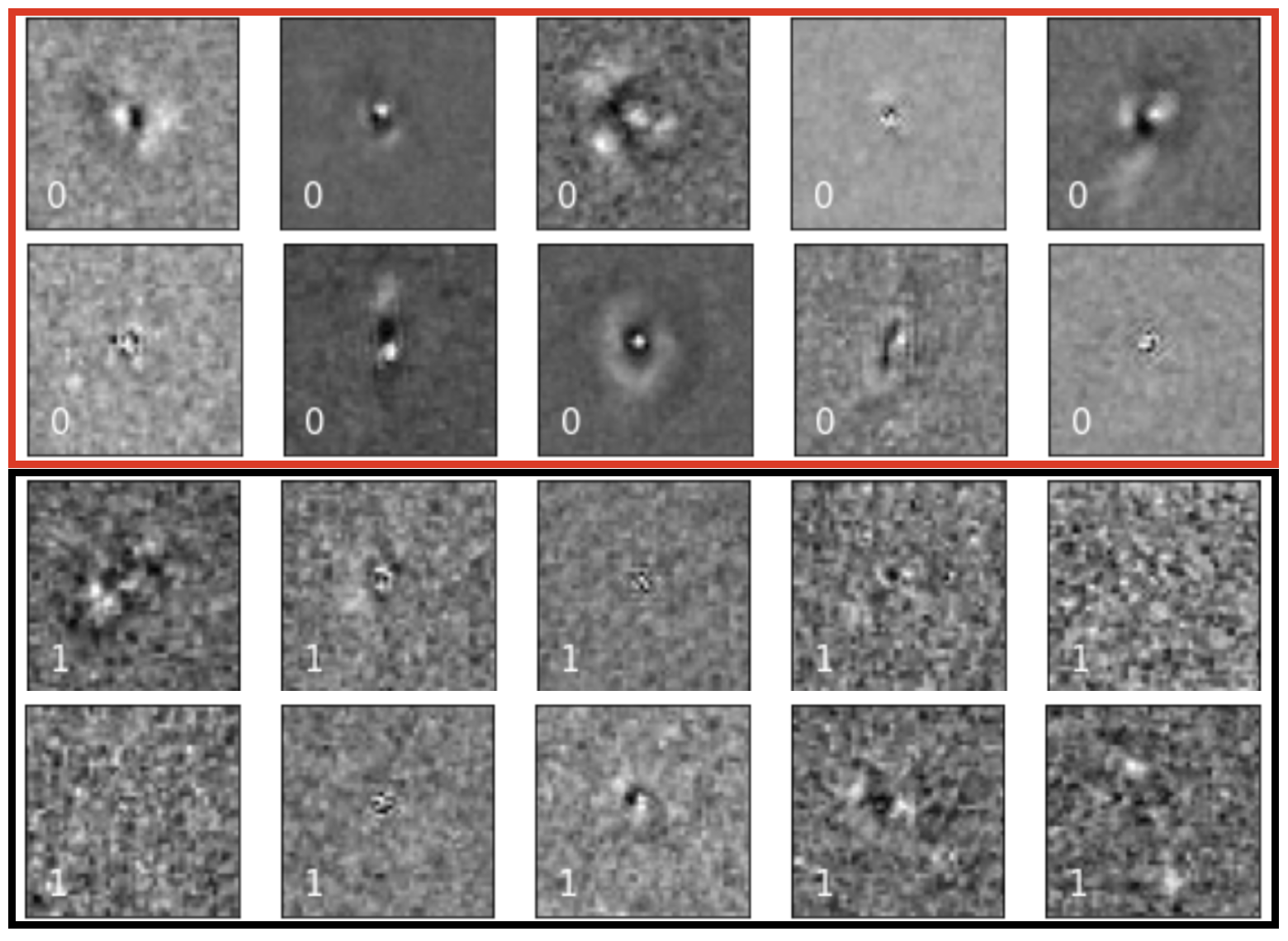}
    \caption[Visualization of residuals in the GMM-clustered unsupervised CvAE latent space ]{Following the similar layout as in Figure\,\ref{fig:supervise_gmm_res}, we show the residual images  sampled randomly from the GMM-based clusters identified in the unsupervised CvAE based PCA space.}
    \label{fig:unsupervise_gmm_res}
\end{figure}

Following a similar approach to our supervised case, we run an initial $k$-means inertia test for our unsupervised CvAE latent features in PCA space, and find that it allows an optimal $k=2$ clusters (see Figure\,\ref{fig:unsupervise_pca_gmm}. Using this information, we run GMM-based clustering to identify $2$ clusters in the CvAE-based PCA space (see right panel in Figure\,\ref{fig:unsupervise_pca_gmm}). After running an SVC classifier with a ``Linear'' kernel (owing to the simplistic nature of the PC1 vs PC2 distribution), we find that these two GMM-based clusters ($n_{\rm gmm,vae}$) are split roughly at $PC1\sim 0$, where $n_{\rm gmm,vae} = 1$ ($= 0$) cluster quantitatively corresponds to objects with smaller (larger) $SPF$ values, respectively (see Figure\,\ref{fig:unsupervised_pca_SPF_B_RFF}). In Figure\,\ref{fig:unsupervise_gmm_res}, we visually show example residual images randomly sampled per GMM-identified cluster. We find that the $n_{\rm gmm,vae} =0$ residuals host a mixture of visually strong and intermediate strength residuals, and the $n_{\rm gmm,vae} =1$ residuals are mostly ``clean'' with some hosting weak diffuse signatures. This exercise reaffirms our previous discussion that the unsupervised latent features are less informative and lack clear discriminatory power in terms of characterizing residual substructure.

\section{Conclusions} \label{sec:conclusions}
In this paper, we carryout a Deep Learning (DL) based exploratory study of the residual substructures hosted by a large sample of $\sim 10,000$ massive and bright galaxies from the {\it Hubble Space Telescope} ({\it HST}) CANDELS survey spanning a redshift range $1<z<3$. We develop supervised Convolutional Neural Network (CNN) and unsupervised Convolutional Variational Autoencoder (CvAE) frameworks to primarily extract the deep latent space features representing the different residual substructure characteristics. We analyze the latent space using Principle Component Analysis in conjunction with independently quantified metrics of residual strength. We assess the DL-based latent features' ability to distinguish different residual characteristics by using  Gaussian Mixture Modeling (GMM) based unsupervised clustering and Support Vector Classifiers (SVC). The key methodological steps and their corresponding results are as follows:
\begin{itemize}
    \item We select a sample of $10,046$ massive galaxies ($M_{\rm stellar} \geq 10^{9.5}\,M_{\odot}$) spanning $1<z<3$ from the CANDELS survey that are also {\it HST} {\it H}-band bright ($H<24.5\,{\rm mag}$) and have good photometric data (PhotFlag$=0$). For this sample, we procure the single-S\'ersic profile-fitting based residual images generated by \cite{van_der_wel_12}, carryout a visual-based characterization of different residual features hosted by them, and compile them into five residual classes -- Clean, Core, Peculiar, Asymmetric, and General. 
    \item We prepare our sample's residual images to serve as inputs to our DL networks by pre-processing them such that the resultant image set comprises of postage-stamp cutouts of the raw residual images which have been modified to contain only the ``galaxy of interest''. We split our main sample into training and testing subsets, and augment our training data to span equi-proportionately across the five residual characteristic classes. Finally, we derive three residual metrics that quantitatively describe their strengths: Significant Pixel Flux ($SPF$), Bumpiness ($B$), and Residual Flux Fraction ($RFF$), which we use in our analysis to assess the latent space features of our DL networks.
    \item We implement a $5$-layer deep CNN network and train it on the augmented training image sample and visual-based residual class labels as targets. We extract the latent space features from the pre-classification fully-connected layer and perform PCA analysis on it. We assess the latent features in the first two principle component axes and find that the Clean vs. Peculiar and General classes are distinctly separable, while the remaining Core and Asymmetric classes lie intermediate to them with considerable (by eye) overlap. 
    \item By correlating the supervised CNN-based latent features in PCA space with the residual quantitative metrics ($SPF$, $B$, and  $RFF$), we find that the $PC1$ eigen axis correlates strongly with $SPF$ and relatively weakly with $B$ and $RFF$, where objects with $PC1\gtrsim 1$ have higher values in all three metrics, and vice-versa.
    \item Informed by a $k$-means clustering inertia test on the supervised CNN-based latent features in PCA space, we identify $6$ clusters ($n_{\rm gmm,cnn}$) using GMM and find separating decision boundaries using SVC. We find that $n_{\rm gmm,cnn} = 2$ group exclusively contains objects hosting strong symmetric and asymmetric residuals, whereas the $n_{\rm gmm,cnn} = 3,4$ clusters host centrally dominant and ``Clean'' residuals, respectively. On the other hand, the clusters $n_{\rm gmm,cnn} = 0,1,5$  comprise weaker diffuse residual signatures. We find that our CNN-based latent space offers promise towards naturally distinguishing different residual characteristics.
    \item We also implement an unsupervised CvAE with $3$-layered Encoder-Decoder framework and train it on our augmented training image data. We extract the learned latent space features from the encoder network and perform PCA on it. We assess this latent space in the prominent PCA eigen axes PC1 vs. PC2 and find that its overall visual-class wise distribution follows a continuum, where the Peculiar and General classes appear as outliers to the Clean distribution, and the Asymmetric and Core classes substantially overlap with the remaining remaining. 
    \item We correlate the unsupervised latent features in PCA space with the quantitative residual metrics ($SPF$, $B$, and  $RFF$) and find that only the $PC1$ eigen axis correlates strongly with the $SPF$ values and weakly with $B$ and $RFF$ values.
    \item Using a $k$-means inertia test, we find that the CvAE-based latent features are best described by utmost $2$ clusters. We apply GMM-based clustering to identify $2$ clusters in the PCA axes and define a decision boundary using a SVC classifier. We  find that the two identified clusters ($n_{\rm gmm,vae}$) are split at $PC1\sim 0$, where the $n_{\rm gmm,vae} = 0$ ($= 1$) cluster hosts a mixture of strong and intermediate strength residuals (clean and weak residuals), respectively. We find that our unsupervised CvAE based latent features although offer some insights, lack clear discriminatory power when characterizing residual substructure.
\end{itemize}
Our methodological framework and the informative results from our supervised and unsupervised DL models offers a promising path towards implementing automated methods for residual feature assessment. These are especially essential in the context of future large-scale telescope surveys.

\section*{Acknowledgments \& Disclosures}
This work was originally inspired during the early times of ``computer vision boom'' and was performed during KBM's doctoral dissertation (2019-2020). However, this work never was able to make it fully to the peer review stage owing to KBM's many other commitments. In the interest of getting these insights to the astronomy community at large, KBM is releasing this {\bf non peer-reviewed} version to the community.      

KBM acknowledges helpful insights and discussions with Marc Huertas-Company, Kartheik Iyer, Viraj Pandya, Marziye Jafariyazani, Erin Kado-Fong, Allison Kirkpatrick, David Koo, Peter Kurczynski, Bret Lehmer, Jennifer Lotz, Bahram Mobasher, and Ripon Saha. KBM, DHM, LDL, and RE acknowledge funding from the NASA Hubble Space Telescope Archival Research grant 15040. 

KBM conveys his sincere thanks to Joel Primack for all the inspiring conversations and support. Rest easy, Joel! 



\bibliographystyle{mnras}
\bibliography{bibliography.bib} 

\begin{thebibliography}{}
\makeatletter
\relax
\def\mn@urlcharsother{\let\do\@makeother \do\$\do\&\do\#\do\^\do\_\do\%\do\~}
\def\mn@doi{\begingroup\mn@urlcharsother \@ifnextchar [ {\mn@doi@} {\mn@doi@[]}}
\def\mn@doi@[#1]#2{\def\@tempa{#1}\ifx\@tempa\@empty \href {http://dx.doi.org/#2} {doi:#2}\else \href {http://dx.doi.org/#2} {#1}\fi \endgroup}
\def\mn@eprint#1#2{\mn@eprint@#1:#2::\@nil}
\def\mn@eprint@arXiv#1{\href {http://arxiv.org/abs/#1} {{\tt arXiv:#1}}}
\def\mn@eprint@dblp#1{\href {http://dblp.uni-trier.de/rec/bibtex/#1.xml} {dblp:#1}}
\def\mn@eprint@#1:#2:#3:#4\@nil{\def\@tempa {#1}\def\@tempb {#2}\def\@tempc {#3}\ifx \@tempc \@empty \let \@tempc \@tempb \let \@tempb \@tempa \fi \ifx \@tempb \@empty \def\@tempb {arXiv}\fi \@ifundefined {mn@eprint@\@tempb}{\@tempb:\@tempc}{\expandafter \expandafter \csname mn@eprint@\@tempb\endcsname \expandafter{\@tempc}}}

\bibitem[\protect\citeauthoryear{{Abraham}, {van den Bergh}, {Glazebrook}, {Ellis}, {Santiago}, {Surma}  \& {Griffiths}}{{Abraham} et~al.}{1996a}]{Abraham96b}
{Abraham} R.~G.,  {van den Bergh} S.,  {Glazebrook} K.,  {Ellis} R.~S.,  {Santiago} B.~X.,  {Surma} P.,   {Griffiths} R.~E.,  1996a, \mn@doi [\apjs] {10.1086/192352}, \href {https://ui.adsabs.harvard.edu/abs/1996ApJS..107....1A} {107, 1}

\bibitem[\protect\citeauthoryear{{Abraham}, {Tanvir}, {Santiago}, {Ellis}, {Glazebrook}  \& {van den Bergh}}{{Abraham} et~al.}{1996b}]{Abraham96a}
{Abraham} R.~G.,  {Tanvir} N.~R.,  {Santiago} B.~X.,  {Ellis} R.~S.,  {Glazebrook} K.,   {van den Bergh} S.,  1996b, \mn@doi [\mnras] {10.1093/mnras/279.3.L47}, \href {https://ui.adsabs.harvard.edu/abs/1996MNRAS.279L..47A} {279, L47}

\bibitem[\protect\citeauthoryear{{Abraham}, {van den Bergh}  \& {Nair}}{{Abraham} et~al.}{2003}]{Abraham03}
{Abraham} R.~G.,  {van den Bergh} S.,   {Nair} P.,  2003, \mn@doi [\apj] {10.1086/373919}, \href {http://adsabs.harvard.edu/abs/2003ApJ...588..218A} {588, 218}

\bibitem[\protect\citeauthoryear{Alpert, Kahng  \& Yao}{Alpert et~al.}{1999}]{alpert99}
Alpert C.~J.,  Kahng A.~B.,   Yao S.-Z.,  1999, Discrete Applied Mathematics, 90, 3

\bibitem[\protect\citeauthoryear{Barbary, Boone, McCully, Craig, Deil  \& Rose}{Barbary et~al.}{2016}]{barbary16}
Barbary K.,  Boone K.,  McCully C.,  Craig M.,  Deil C.,   Rose B.,  2016, kbarbary/sep: v1.0.0, \mn@doi{10.5281/zenodo.159035}, \url {https://doi.org/10.5281/zenodo.159035}

\bibitem[\protect\citeauthoryear{{Barden}, {H{\"a}u{\ss}ler}, {Peng}, {McIntosh}  \& {Guo}}{{Barden} et~al.}{2012}]{Barden12}
{Barden} M.,  {H{\"a}u{\ss}ler} B.,  {Peng} C.~Y.,  {McIntosh} D.~H.,   {Guo} Y.,  2012, \mn@doi [\mnras] {10.1111/j.1365-2966.2012.20619.x}, \href {http://adsabs.harvard.edu/abs/2012MNRAS.422..449B} {422, 449}

\bibitem[\protect\citeauthoryear{{Barnes} \& {Hernquist}}{{Barnes} \& {Hernquist}}{1996}]{BarnesHernquist96}
{Barnes} J.~E.,  {Hernquist} L.,  1996, \mn@doi [\apj] {10.1086/177957}, \href {http://adsabs.harvard.edu/abs/1996ApJ...471..115B} {471, 115}

\bibitem[\protect\citeauthoryear{{Barro} et~al.,}{{Barro} et~al.}{2019}]{Barro19}
{Barro} G.,  et~al., 2019, \mn@doi [\apjs] {10.3847/1538-4365/ab23f2}, \href {https://ui.adsabs.harvard.edu/abs/2019ApJS..243...22B} {243, 22}

\bibitem[\protect\citeauthoryear{Bell et~al.,}{Bell et~al.}{2006a}]{bell_dry_2006}
Bell E.~F.,  et~al., 2006a, \mn@doi [ApJ] {10.1086/499931}, 640, 241

\bibitem[\protect\citeauthoryear{{Bell}, {Phleps}, {Somerville}, {Wolf}, {Borch}  \& {Meisenheimer}}{{Bell} et~al.}{2006b}]{bell06b}
{Bell} E.~F.,  {Phleps} S.,  {Somerville} R.~S.,  {Wolf} C.,  {Borch} A.,   {Meisenheimer} K.,  2006b, \mn@doi [\apj] {10.1086/508408}, \href {http://adsabs.harvard.edu/abs/2006ApJ...652..270B} {652, 270}

\bibitem[\protect\citeauthoryear{{Bertin} \& {Arnouts}}{{Bertin} \& {Arnouts}}{1996}]{bertin96}
{Bertin} E.,  {Arnouts} S.,  1996, \mn@doi [\aaps] {10.1051/aas:1996164}, \href {http://adsabs.harvard.edu/abs/1996A%26AS..117..393B} {117, 393}

\bibitem[\protect\citeauthoryear{{Blakeslee} et~al.,}{{Blakeslee} et~al.}{2006}]{Blakeslee06}
{Blakeslee} J.~P.,  et~al., 2006, \mn@doi [ApJ] {10.1086/503539}, \href {https://ui.adsabs.harvard.edu/abs/2006ApJ...644...30B} {644, 30}

\bibitem[\protect\citeauthoryear{Boser, Guyon  \& Vapnik}{Boser et~al.}{}]{Boser92}
Boser B.~E.,  Guyon I.~M.,   Vapnik V.~N., , in Proceedings of the 5th Annual ACM Workshop on Computational Learning Theory. pp 144--152

\bibitem[\protect\citeauthoryear{{Bournaud}, {Dekel}, {Teyssier}, {Cacciato}, {Daddi}, {Juneau}  \& {Shankar}}{{Bournaud} et~al.}{2011}]{Bournaud11}
{Bournaud} F.,  {Dekel} A.,  {Teyssier} R.,  {Cacciato} M.,  {Daddi} E.,  {Juneau} S.,   {Shankar} F.,  2011, \mn@doi [\apjl] {10.1088/2041-8205/741/2/L33}, \href {http://adsabs.harvard.edu/abs/2011ApJ...741L..33B} {741, L33}

\bibitem[\protect\citeauthoryear{{Cassata} et~al.,}{{Cassata} et~al.}{2005}]{cassata05}
{Cassata} P.,  et~al., 2005, \mn@doi [\mnras] {10.1111/j.1365-2966.2005.08657.x}, \href {http://adsabs.harvard.edu/abs/2005MNRAS.357..903C} {357, 903}

\bibitem[\protect\citeauthoryear{{Ceverino}, {Dekel}, {Tweed}  \& {Primack}}{{Ceverino} et~al.}{2015}]{Ceverino15a}
{Ceverino} D.,  {Dekel} A.,  {Tweed} D.,   {Primack} J.,  2015, \mn@doi [\mnras] {10.1093/mnras/stu2694}, \href {https://ui.adsabs.harvard.edu/abs/2015MNRAS.447.3291C} {447, 3291}

\bibitem[\protect\citeauthoryear{{Cheng}, {Li}, {Conselice}, {Arag{\'o}n-Salamanca}, {Dye}  \& {Metcalf}}{{Cheng} et~al.}{2020}]{Cheng20}
{Cheng} T.-Y.,  {Li} N.,  {Conselice} C.~J.,  {Arag{\'o}n-Salamanca} A.,  {Dye} S.,   {Metcalf} R.~B.,  2020, \mn@doi [\mnras] {10.1093/mnras/staa1015}, \href {https://ui.adsabs.harvard.edu/abs/2020MNRAS.494.3750C} {494, 3750}

\bibitem[\protect\citeauthoryear{{Conselice}}{{Conselice}}{2003}]{conselice_03_cas}
{Conselice} C.~J.,  2003, \mn@doi [\apjs] {10.1086/375001}, \href {http://adsabs.harvard.edu/abs/2003ApJS..147....1C} {147, 1}

\bibitem[\protect\citeauthoryear{{Conselice}, {Bershady}, {Dickinson}  \& {Papovich}}{{Conselice} et~al.}{2003}]{conselice03a}
{Conselice} C.~J.,  {Bershady} M.~A.,  {Dickinson} M.,   {Papovich} C.,  2003, \mn@doi [\aj] {10.1086/377318}, \href {http://adsabs.harvard.edu/abs/2003AJ....126.1183C} {126, 1183}

\bibitem[\protect\citeauthoryear{Conselice, Rajgor  \& Myers}{Conselice et~al.}{2008}]{conselice_structures_2008}
Conselice C.~J.,  Rajgor S.,   Myers R.,  2008, \mn@doi [Monthly Notices of the Royal Astronomical Society] {10.1111/j.1365-2966.2008.13069.x}, 386, 909

\bibitem[\protect\citeauthoryear{Conselice, Yang  \& Bluck}{Conselice et~al.}{2009}]{conselice_structures_2009}
Conselice C.~J.,  Yang C.,   Bluck A. F.~L.,  2009, \mn@doi [Monthly Notices of the Royal Astronomical Society] {10.1111/j.1365-2966.2009.14396.x}, 394, 1956

\bibitem[\protect\citeauthoryear{{Cox}, {Jonsson}, {Somerville}, {Primack}  \& {Dekel}}{{Cox} et~al.}{2008}]{Cox08}
{Cox} T.~J.,  {Jonsson} P.,  {Somerville} R.~S.,  {Primack} J.~R.,   {Dekel} A.,  2008, \mn@doi [\mnras] {10.1111/j.1365-2966.2007.12730.x}, \href {http://adsabs.harvard.edu/abs/2008MNRAS.384..386C} {384, 386}

\bibitem[\protect\citeauthoryear{Dahlen et~al.,}{Dahlen et~al.}{2013}]{dahlen_critical_2013}
Dahlen T.,  et~al., 2013, \mn@doi [ApJ] {10.1088/0004-637X/775/2/93}, 775, 93

\bibitem[\protect\citeauthoryear{{Dekel} \& {Burkert}}{{Dekel} \& {Burkert}}{2014}]{Dekel14}
{Dekel} A.,  {Burkert} A.,  2014, \mn@doi [\mnras] {10.1093/mnras/stt2331}, \href {http://adsabs.harvard.edu/abs/2014MNRAS.438.1870D} {438, 1870}

\bibitem[\protect\citeauthoryear{{Di Matteo}, {Combes}, {Melchior}  \& {Semelin}}{{Di Matteo} et~al.}{2007}]{Dimatteo07}
{Di Matteo} P.,  {Combes} F.,  {Melchior} A.-L.,   {Semelin} B.,  2007, \mn@doi [\aap] {10.1051/0004-6361:20066959}, \href {http://adsabs.harvard.edu/abs/2007A%26A...468...61D} {468, 61}

\bibitem[\protect\citeauthoryear{{Di Matteo}, {Bournaud}, {Martig}, {Combes}, {Melchior}  \& {Semelin}}{{Di Matteo} et~al.}{2008}]{DiMatteo08}
{Di Matteo} P.,  {Bournaud} F.,  {Martig} M.,  {Combes} F.,  {Melchior} A.-L.,   {Semelin} B.,  2008, \mn@doi [\aap] {10.1051/0004-6361:200809480}, \href {http://adsabs.harvard.edu/abs/2008A%26A...492...31D} {492, 31}

\bibitem[\protect\citeauthoryear{{Dieleman}, {Willett}  \& {Dambre}}{{Dieleman} et~al.}{2015}]{Dieleman15}
{Dieleman} S.,  {Willett} K.~W.,   {Dambre} J.,  2015, \mn@doi [\mnras] {10.1093/mnras/stv632}, \href {https://ui.adsabs.harvard.edu/abs/2015MNRAS.450.1441D} {450, 1441}

\bibitem[\protect\citeauthoryear{{Dom{\'\i}nguez S{\'a}nchez}, {Huertas-Company}, {Bernardi}, {Tuccillo}  \& {Fischer}}{{Dom{\'\i}nguez S{\'a}nchez} et~al.}{2018}]{Dominguez-Sanchez18}
{Dom{\'\i}nguez S{\'a}nchez} H.,  {Huertas-Company} M.,  {Bernardi} M.,  {Tuccillo} D.,   {Fischer} J.~L.,  2018, \mn@doi [\mnras] {10.1093/mnras/sty338}, \href {https://ui.adsabs.harvard.edu/abs/2018MNRAS.476.3661D} {476, 3661}

\bibitem[\protect\citeauthoryear{{Duc} \& {Renaud}}{{Duc} \& {Renaud}}{2013}]{Duc2013}
{Duc} P.-A.,  {Renaud} F.,  2013, in {Souchay} J.,  {Mathis} S.,   {Tokieda} T.,  eds,  Lecture Notes in Physics, Berlin Springer Verlag Vol. 861, Lecture Notes in Physics, Berlin Springer Verlag. p.~327 (\mn@eprint {arXiv} {1112.1922}), \mn@doi{10.1007/978-3-642-32961-6_9}

\bibitem[\protect\citeauthoryear{{Duncan} et~al.,}{{Duncan} et~al.}{2019}]{Duncan19}
{Duncan} K.,  et~al., 2019, \mn@doi [\apj] {10.3847/1538-4357/ab148a}, \href {https://ui.adsabs.harvard.edu/abs/2019ApJ...876..110D} {876, 110}

\bibitem[\protect\citeauthoryear{{Eneev}, {Kozlov}  \& {Sunyaev}}{{Eneev} et~al.}{1973}]{Eneev73}
{Eneev} T.~M.,  {Kozlov} N.~N.,   {Sunyaev} R.~A.,  1973, \aap, \href {http://adsabs.harvard.edu/abs/1973A%26A....22...41E} {22, 41}

\bibitem[\protect\citeauthoryear{Ester, Kriegel, Sander  \& Xu}{Ester et~al.}{1996}]{Ester96}
Ester M.,  Kriegel H.-P.,  Sander J.,   Xu X.,  1996, in Proceedings of the Second International Conference on Knowledge Discovery and Data Mining. KDD'96.
AAAI Press, p. 226–231

\bibitem[\protect\citeauthoryear{{Ferreira}, {Conselice}, {Duncan}, {Cheng}, {Griffiths}  \& {Whitney}}{{Ferreira} et~al.}{2020}]{Ferreira20}
{Ferreira} L.,  {Conselice} C.~J.,  {Duncan} K.,  {Cheng} T.-Y.,  {Griffiths} A.,   {Whitney} A.,  2020, \mn@doi [\apj] {10.3847/1538-4357/ab8f9b}, \href {https://ui.adsabs.harvard.edu/abs/2020ApJ...895..115F} {895, 115}

\bibitem[\protect\citeauthoryear{Fukushima \& Miyake}{Fukushima \& Miyake}{1982}]{Fukushima82}
Fukushima K.,  Miyake S.,  1982, in Amari S.-i.,  Arbib M.~A.,  eds, Competition and Cooperation in Neural Nets. Springer Berlin Heidelberg, Berlin, Heidelberg, pp 267--285

\bibitem[\protect\citeauthoryear{Galametz et~al.,}{Galametz et~al.}{2013}]{galametz_candels_2013}
Galametz A.,  et~al., 2013, \mn@doi [ApJ Supplement Series] {10.1088/0067-0049/206/2/10}, 206, 10

\bibitem[\protect\citeauthoryear{Grogin et~al.,}{Grogin et~al.}{2011}]{grogin_candels:_2011}
Grogin N.~A.,  et~al., 2011, \mn@doi [ApJ Supplement Series] {10.1088/0067-0049/197/2/35}, 197, 35

\bibitem[\protect\citeauthoryear{Guo et~al.,}{Guo et~al.}{2013}]{guo_candels_2013}
Guo Y.,  et~al., 2013, \mn@doi [ApJ Supplement Series] {10.1088/0067-0049/207/2/24}, 207, 24

\bibitem[\protect\citeauthoryear{Hearst}{Hearst}{1998}]{Hearst98}
Hearst M.~A.,  1998, \mn@doi [IEEE Intelligent Systems] {10.1109/5254.708428}, 13, 18–28

\bibitem[\protect\citeauthoryear{{Hocking}, {Geach}, {Sun}  \& {Davey}}{{Hocking} et~al.}{2018}]{Hocking18}
{Hocking} A.,  {Geach} J.~E.,  {Sun} Y.,   {Davey} N.,  2018, \mn@doi [\mnras] {10.1093/mnras/stx2351}, \href {http://adsabs.harvard.edu/abs/2018MNRAS.473.1108H} {473, 1108}

\bibitem[\protect\citeauthoryear{{Hopkins}, {Somerville}, {Hernquist}, {Cox}, {Robertson}  \& {Li}}{{Hopkins} et~al.}{2006}]{Hopkins06a}
{Hopkins} P.~F.,  {Somerville} R.~S.,  {Hernquist} L.,  {Cox} T.~J.,  {Robertson} B.,   {Li} Y.,  2006, \mn@doi [\apj] {10.1086/508503}, \href {http://adsabs.harvard.edu/abs/2006ApJ...652..864H} {652, 864}

\bibitem[\protect\citeauthoryear{Hopkins et~al.,}{Hopkins et~al.}{2010}]{hopkins_mergers_2010}
Hopkins P.~F.,  et~al., 2010, \mn@doi [ApJ] {10.1088/0004-637X/715/1/202}, 715, 202

\bibitem[\protect\citeauthoryear{{Hoyos} et~al.,}{{Hoyos} et~al.}{2012}]{Hoyos12}
{Hoyos} C.,  et~al., 2012, \mn@doi [\mnras] {10.1111/j.1365-2966.2011.19918.x}, \href {http://adsabs.harvard.edu/abs/2012MNRAS.419.2703H} {419, 2703}

\bibitem[\protect\citeauthoryear{{Huertas-Company}, {Aguerri}, {Bernardi}, {Mei}  \& {S{\'a}nchez Almeida}}{{Huertas-Company} et~al.}{2011}]{Huertas-Company11}
{Huertas-Company} M.,  {Aguerri} J.~A.~L.,  {Bernardi} M.,  {Mei} S.,   {S{\'a}nchez Almeida} J.,  2011, \mn@doi [\aap] {10.1051/0004-6361/201015735}, \href {https://ui.adsabs.harvard.edu/abs/2011A&A...525A.157H} {525, A157}

\bibitem[\protect\citeauthoryear{{Huertas-Company} et~al.,}{{Huertas-Company} et~al.}{2015}]{Huertas-Company15}
{Huertas-Company} M.,  et~al., 2015, \mn@doi [\apjs] {10.1088/0067-0049/221/1/8}, \href {https://ui.adsabs.harvard.edu/abs/2015ApJS..221....8H} {221, 8}

\bibitem[\protect\citeauthoryear{{Huertas-Company} et~al.,}{{Huertas-Company} et~al.}{2018}]{Huertas-Company18}
{Huertas-Company} M.,  et~al., 2018, preprint, \href {http://adsabs.harvard.edu/abs/2018arXiv180407307H} {} (\mn@eprint {arXiv} {1804.07307})

\bibitem[\protect\citeauthoryear{{Huertas-Company} et~al.,}{{Huertas-Company} et~al.}{2019}]{Huertas-Company19}
{Huertas-Company} M.,  et~al., 2019, \mn@doi [\mnras] {10.1093/mnras/stz2191}, \href {https://ui.adsabs.harvard.edu/abs/2019MNRAS.489.1859H} {489, 1859}

\bibitem[\protect\citeauthoryear{{Jogee}}{{Jogee}}{2009}]{Jogee09}
{Jogee} S.,  2009, in {Andersen} J.,  {Nordstr{\"o}ara} {m} B.,   {Bland-Hawthorn} J.,  eds,  IAU Symposium Vol. 254, The Galaxy Disk in Cosmological Context. pp 67--72 (\mn@eprint {arXiv} {0810.5617}), \mn@doi{10.1017/S1743921308027403}

\bibitem[\protect\citeauthoryear{Jolliffe}{Jolliffe}{1986}]{Jolliffe86}
Jolliffe I.~T.,  1986, Principal Components in Regression Analysis.
Springer New York, New York, NY, pp 129--155, \mn@doi{10.1007/978-1-4757-1904-8_8}, \url {https://doi.org/10.1007/978-1-4757-1904-8_8}

\bibitem[\protect\citeauthoryear{Kingma \& Ba}{Kingma \& Ba}{2014}]{kingma17}
Kingma D.~P.,  Ba J.,  2014, arXiv preprint arXiv:1412.6980

\bibitem[\protect\citeauthoryear{Kocevski et~al.,}{Kocevski et~al.}{2012}]{kocevski_candels:_2012}
Kocevski D.~D.,  et~al., 2012, \mn@doi [ApJ] {10.1088/0004-637X/744/2/148}, 744, 148

\bibitem[\protect\citeauthoryear{{Kodra}}{{Kodra}}{2019}]{Kodra19}
{Kodra} D.,  2019, PhD thesis, University of Pittsburgh

\bibitem[\protect\citeauthoryear{{Koekemoer} et~al.,}{{Koekemoer} et~al.}{2011}]{Koekemoer}
{Koekemoer} A.~M.,  et~al., 2011, \mn@doi [\apjs] {10.1088/0067-0049/197/2/36}, \href {http://adsabs.harvard.edu/abs/2011ApJS..197...36K} {197, 36}

\bibitem[\protect\citeauthoryear{{Lackner} et~al.,}{{Lackner} et~al.}{2014}]{Lackner14}
{Lackner} C.~N.,  et~al., 2014, \mn@doi [\aj] {10.1088/0004-6256/148/6/137}, \href {http://adsabs.harvard.edu/abs/2014AJ....148..137L} {148, 137}

\bibitem[\protect\citeauthoryear{Lecun, Bottou, Bengio  \& Haffner}{Lecun et~al.}{1998}]{Lecun98}
Lecun Y.,  Bottou L.,  Bengio Y.,   Haffner P.,  1998, \mn@doi [Proceedings of the IEEE] {10.1109/5.726791}, 86, 2278

\bibitem[\protect\citeauthoryear{{Lin} et~al.,}{{Lin} et~al.}{2004}]{Lin04}
{Lin} L.,  et~al., 2004, \mn@doi [\apjl] {10.1086/427183}, \href {http://adsabs.harvard.edu/abs/2004ApJ...617L...9L} {617, L9}

\bibitem[\protect\citeauthoryear{Lloyd}{Lloyd}{1957}]{lloyd57}
Lloyd S.,  1957, IEEE Trans. Inform. Theor.(1957/1982), 18

\bibitem[\protect\citeauthoryear{L\'opez-Sanjuan, Balcells, Pérez-González, Barro, García-Dabó, Gallego  \& Zamorano}{L\'opez-Sanjuan et~al.}{2009}]{lopez-sanjuan_galaxy_2009}
L\'opez-Sanjuan C.,  Balcells M.,  Pérez-González P.~G.,  Barro G.,  García-Dabó C.~E.,  Gallego J.,   Zamorano J.,  2009, \mn@doi [Astronomy and Astrophysics] {10.1051/0004-6361/200911923}, 501, 505

\bibitem[\protect\citeauthoryear{{Lotz} et~al.,}{{Lotz} et~al.}{2008}]{lotz08}
{Lotz} J.~M.,  et~al., 2008, \mn@doi [\apj] {10.1086/523659}, \href {http://adsabs.harvard.edu/abs/2008ApJ...672..177L} {672, 177}

\bibitem[\protect\citeauthoryear{Lotz, Jonsson, Cox, Croton, Primack, Somerville  \& Stewart}{Lotz et~al.}{2011}]{lotz_major_2011}
Lotz J.~M.,  Jonsson P.,  Cox T.~J.,  Croton D.,  Primack J.~R.,  Somerville R.~S.,   Stewart K.,  2011, \mn@doi [ApJ] {10.1088/0004-637X/742/2/103}, 742, 103

\bibitem[\protect\citeauthoryear{MacQueen et~al.}{MacQueen et~al.}{1967}]{macqueen67}
MacQueen J.,  et~al., 1967, in Proceedings of the fifth Berkeley symposium on mathematical statistics and probability. pp 281--297

\bibitem[\protect\citeauthoryear{{Man}, {Zirm}  \& {Toft}}{{Man} et~al.}{2016}]{man16}
{Man} A.~W.~S.,  {Zirm} A.~W.,   {Toft} S.,  2016, \mn@doi [\apj] {10.3847/0004-637X/830/2/89}, \href {http://adsabs.harvard.edu/abs/2016ApJ...830...89M} {830, 89}

\bibitem[\protect\citeauthoryear{{Mantha} et~al.,}{{Mantha} et~al.}{2018}]{Mantha18}
{Mantha} K.~B.,  et~al., 2018, \mn@doi [\mnras] {10.1093/mnras/stx3260}, \href {http://adsabs.harvard.edu/abs/2018MNRAS.475.1549M} {475, 1549}

\bibitem[\protect\citeauthoryear{{Mantha} et~al.,}{{Mantha} et~al.}{2019}]{Mantha19}
{Mantha} K.~B.,  et~al., 2019, \mn@doi [\mnras] {10.1093/mnras/stz872}, \href {https://ui.adsabs.harvard.edu/abs/2019MNRAS.486.2643M} {486, 2643}

\bibitem[\protect\citeauthoryear{Masci, Meier, Cire{\c{s}}an  \& Schmidhuber}{Masci et~al.}{2011}]{Masci11}
Masci J.,  Meier U.,  Cire{\c{s}}an D.,   Schmidhuber J.,  2011, in Honkela T.,  Duch W.,  Girolami M.,   Kaski S.,  eds, Artificial Neural Networks and Machine Learning -- ICANN 2011. Springer Berlin Heidelberg, Berlin, Heidelberg, pp 52--59

\bibitem[\protect\citeauthoryear{{McIntosh}, {Guo}, {Hertzberg}, {Katz}, {Mo}, {van den Bosch}  \& {Yang}}{{McIntosh} et~al.}{2008}]{mcintosh_2008}
{McIntosh} D.~H.,  {Guo} Y.,  {Hertzberg} J.,  {Katz} N.,  {Mo} H.~J.,  {van den Bosch} F.~C.,   {Yang} X.,  2008, \mn@doi [\mnras] {10.1111/j.1365-2966.2008.13531.x}, \href {http://adsabs.harvard.edu/abs/2008MNRAS.388.1537M} {388, 1537}

\bibitem[\protect\citeauthoryear{McLachlan \& Basford}{McLachlan \& Basford}{1988}]{mclachlan88}
McLachlan G.~J.,  Basford K.~E.,  1988, Mixture models: Inference and applications to clustering.
 Vol. 38, M. Dekker New York

\bibitem[\protect\citeauthoryear{{Metcalf} et~al.,}{{Metcalf} et~al.}{2019}]{Metcalf19}
{Metcalf} R.~B.,  et~al., 2019, \mn@doi [\aap] {10.1051/0004-6361/201832797}, \href {https://ui.adsabs.harvard.edu/abs/2019A&A...625A.119M} {625, A119}

\bibitem[\protect\citeauthoryear{{Mobasher} et~al.,}{{Mobasher} et~al.}{2015}]{Mobasher15}
{Mobasher} B.,  et~al., 2015, \mn@doi [\apj] {10.1088/0004-637X/808/1/101}, \href {http://adsabs.harvard.edu/abs/2015ApJ...808..101M} {808, 101}

\bibitem[\protect\citeauthoryear{{Mundy}, {Conselice}, {Duncan}, {Almaini}, {H{\"a}u{\ss}ler}  \& {Hartley}}{{Mundy} et~al.}{2017}]{Mundy17}
{Mundy} C.~J.,  {Conselice} C.~J.,  {Duncan} K.~J.,  {Almaini} O.,  {H{\"a}u{\ss}ler} B.,   {Hartley} W.~G.,  2017, preprint, \href {http://adsabs.harvard.edu/abs/2017arXiv170507986M} {} (\mn@eprint {arXiv} {1705.07986})

\bibitem[\protect\citeauthoryear{{Naab}, {Khochfar}  \& {Burkert}}{{Naab} et~al.}{2006}]{Naab06}
{Naab} T.,  {Khochfar} S.,   {Burkert} A.,  2006, \mn@doi [\apjl] {10.1086/500205}, \href {http://adsabs.harvard.edu/abs/2006ApJ...636L..81N} {636, L81}

\bibitem[\protect\citeauthoryear{{Narayanan} et~al.,}{{Narayanan} et~al.}{2010}]{Narayanan10}
{Narayanan} D.,  et~al., 2010, \mn@doi [\mnras] {10.1111/j.1365-2966.2010.16997.x}, \href {http://adsabs.harvard.edu/abs/2010MNRAS.407.1701N} {407, 1701}

\bibitem[\protect\citeauthoryear{{Nayyeri} et~al.,}{{Nayyeri} et~al.}{2017}]{Nayyeri17}
{Nayyeri} H.,  et~al., 2017, \mn@doi [\apjs] {10.3847/1538-4365/228/1/7}, \href {http://adsabs.harvard.edu/abs/2017ApJS..228....7N} {228, 7}

\bibitem[\protect\citeauthoryear{{Patton}, {Pritchet}, {Yee}, {Ellingson}  \& {Carlberg}}{{Patton} et~al.}{1997}]{Patton97}
{Patton} D.~R.,  {Pritchet} C.~J.,  {Yee} H.~K.~C.,  {Ellingson} E.,   {Carlberg} R.~G.,  1997, \apj, \href {http://adsabs.harvard.edu/abs/1997ApJ...475...29P} {475, 29}

\bibitem[\protect\citeauthoryear{Pedregosa et~al.,}{Pedregosa et~al.}{2011}]{scikit-learn}
Pedregosa F.,  et~al., 2011, Journal of Machine Learning Research, 12, 2825

\bibitem[\protect\citeauthoryear{{Peng}, {Ho}, {Impey}  \& {Rix}}{{Peng} et~al.}{2002}]{Peng02}
{Peng} C.~Y.,  {Ho} L.~C.,  {Impey} C.~D.,   {Rix} H.-W.,  2002, \mn@doi [\aj] {10.1086/340952}, \href {http://adsabs.harvard.edu/abs/2002AJ....124..266P} {124, 266}

\bibitem[\protect\citeauthoryear{{Robaina} et~al.,}{{Robaina} et~al.}{2009}]{robaina09}
{Robaina} A.~R.,  et~al., 2009, \mn@doi [\apj] {10.1088/0004-637X/704/1/324}, \href {http://adsabs.harvard.edu/abs/2009ApJ...704..324R} {704, 324}

\bibitem[\protect\citeauthoryear{{Robotham} et~al.,}{{Robotham} et~al.}{2014}]{Robotham14}
{Robotham} A.~S.~G.,  et~al., 2014, \mn@doi [\mnras] {10.1093/mnras/stu1604}, \href {http://adsabs.harvard.edu/abs/2014MNRAS.444.3986R} {444, 3986}

\bibitem[\protect\citeauthoryear{{Sanders}, {Soifer}, {Elias}, {Madore}, {Matthews}, {Neugebauer}  \& {Scoville}}{{Sanders} et~al.}{1988}]{Sanders88a}
{Sanders} D.~B.,  {Soifer} B.~T.,  {Elias} J.~H.,  {Madore} B.~F.,  {Matthews} K.,  {Neugebauer} G.,   {Scoville} N.~Z.,  1988, \mn@doi [\apj] {10.1086/165983}, \href {http://adsabs.harvard.edu/abs/1988ApJ...325...74S} {325, 74}

\bibitem[\protect\citeauthoryear{Santini et~al.,}{Santini et~al.}{2015}]{santini_stellar_2015}
Santini P.,  et~al., 2015, \mn@doi [ApJ] {10.1088/0004-637X/801/2/97}, 801, 97

\bibitem[\protect\citeauthoryear{{Springel}}{{Springel}}{2000}]{Springel00}
{Springel} V.,  2000, \mn@doi [\mnras] {10.1046/j.1365-8711.2000.03187.x}, \href {http://adsabs.harvard.edu/abs/2000MNRAS.312..859S} {312, 859}

\bibitem[\protect\citeauthoryear{{Stefanon} et~al.,}{{Stefanon} et~al.}{2017}]{Stefanon17}
{Stefanon} M.,  et~al., 2017, \mn@doi [\apjs] {10.3847/1538-4365/aa66cb}, \href {http://adsabs.harvard.edu/abs/2017ApJS..229...32S} {229, 32}

\bibitem[\protect\citeauthoryear{{Storey-Fisher}, {Huertas-Company}, {Ramachandra}, {Lanusse}, {Leauthaud}, {Luo}, {Huang}  \& {Prochaska}}{{Storey-Fisher} et~al.}{2021}]{Storey-Fisher21}
{Storey-Fisher} K.,  {Huertas-Company} M.,  {Ramachandra} N.,  {Lanusse} F.,  {Leauthaud} A.,  {Luo} Y.,  {Huang} S.,   {Prochaska} J.~X.,  2021, arXiv e-prints, \href {https://ui.adsabs.harvard.edu/abs/2021arXiv210502434S} {p. arXiv:2105.02434}

\bibitem[\protect\citeauthoryear{{Swinbank} et~al.,}{{Swinbank} et~al.}{2010}]{Swinbank10}
{Swinbank} A.~M.,  et~al., 2010, \mn@doi [\mnras] {10.1111/j.1365-2966.2010.16485.x}, \href {http://adsabs.harvard.edu/abs/2010MNRAS.405..234S} {405, 234}

\bibitem[\protect\citeauthoryear{{Tal} et~al.,}{{Tal} et~al.}{2014}]{Tal_14}
{Tal} T.,  et~al., 2014, \mn@doi [\apj] {10.1088/0004-637X/789/2/164}, \href {http://adsabs.harvard.edu/abs/2014ApJ...789..164T} {789, 164}

\bibitem[\protect\citeauthoryear{{Tohill}, {Ferreira}, {Conselice}, {Bamford}  \& {Ferrari}}{{Tohill} et~al.}{2021}]{Tohill21}
{Tohill} C.,  {Ferreira} L.,  {Conselice} C.~J.,  {Bamford} S.~P.,   {Ferrari} F.,  2021, \mn@doi [\apj] {10.3847/1538-4357/ac033c}, \href {https://ui.adsabs.harvard.edu/abs/2021ApJ...916....4T} {916, 4}

\bibitem[\protect\citeauthoryear{{Toomre} \& {Toomre}}{{Toomre} \& {Toomre}}{1972}]{Toomre72}
{Toomre} A.,  {Toomre} J.,  1972, \mn@doi [\apj] {10.1086/151823}, \href {http://adsabs.harvard.edu/abs/1972ApJ...178..623T} {178, 623}

\bibitem[\protect\citeauthoryear{{Treister}, {Schawinski}, {Urry}  \& {Simmons}}{{Treister} et~al.}{2012}]{Treister12}
{Treister} E.,  {Schawinski} K.,  {Urry} C.~M.,   {Simmons} B.~D.,  2012, \mn@doi [\apjl] {10.1088/2041-8205/758/2/L39}, \href {http://adsabs.harvard.edu/abs/2012ApJ...758L..39T} {758, L39}

\bibitem[\protect\citeauthoryear{{Ventou} et~al.,}{{Ventou} et~al.}{2017}]{Ventou17}
{Ventou} E.,  et~al., 2017, \mn@doi [\aap] {10.1051/0004-6361/201731586}, \href {http://adsabs.harvard.edu/abs/2017A%26A...608A...9V} {608, A9}

\bibitem[\protect\citeauthoryear{{Villforth} et~al.,}{{Villforth} et~al.}{2017}]{villforth17}
{Villforth} C.,  et~al., 2017, \mn@doi [\mnras] {10.1093/mnras/stw3037}, \href {http://adsabs.harvard.edu/abs/2017MNRAS.466..812V} {466, 812}

\bibitem[\protect\citeauthoryear{{Walmsley} et~al.,}{{Walmsley} et~al.}{2020}]{Walmsley20}
{Walmsley} M.,  et~al., 2020, \mn@doi [\mnras] {10.1093/mnras/stz2816}, \href {https://ui.adsabs.harvard.edu/abs/2020MNRAS.491.1554W} {491, 1554}

\bibitem[\protect\citeauthoryear{{Weston}, {McIntosh}, {Brodwin}, {Mann}, {Cooper}, {McConnell}  \& {Nielsen}}{{Weston} et~al.}{2017}]{weston17}
{Weston} M.~E.,  {McIntosh} D.~H.,  {Brodwin} M.,  {Mann} J.,  {Cooper} A.,  {McConnell} A.,   {Nielsen} J.~L.,  2017, \mn@doi [\mnras] {10.1093/mnras/stw2620}, \href {http://adsabs.harvard.edu/abs/2017MNRAS.464.3882W} {464, 3882}

\bibitem[\protect\citeauthoryear{Williams, Quadri  \& Franx}{Williams et~al.}{2011}]{williams_diminishing_2011}
Williams R.~J.,  Quadri R.~F.,   Franx M.,  2011, \mn@doi [ApJ] {10.1088/2041-8205/738/2/L25}, 738, L25

\bibitem[\protect\citeauthoryear{{Wolf} et~al.,}{{Wolf} et~al.}{2005}]{Wolf05}
{Wolf} C.,  et~al., 2005, \mn@doi [\apj] {10.1086/431659}, \href {http://adsabs.harvard.edu/abs/2005ApJ...630..771W} {630, 771}

\bibitem[\protect\citeauthoryear{{Younger}, {Hayward}, {Narayanan}, {Cox}, {Hernquist}  \& {Jonsson}}{{Younger} et~al.}{2009}]{Younger09}
{Younger} J.~D.,  {Hayward} C.~C.,  {Narayanan} D.,  {Cox} T.~J.,  {Hernquist} L.,   {Jonsson} P.,  2009, \mn@doi [\mnras] {10.1111/j.1745-3933.2009.00663.x}, \href {http://adsabs.harvard.edu/abs/2009MNRAS.396L..66Y} {396, L66}

\bibitem[\protect\citeauthoryear{{Zepf} \& {Koo}}{{Zepf} \& {Koo}}{1989}]{Zepf89}
{Zepf} S.~E.,  {Koo} D.~C.,  1989, \mn@doi [\apj] {10.1086/167085}, \href {http://adsabs.harvard.edu/abs/1989ApJ...337...34Z} {337, 34}

\bibitem[\protect\citeauthoryear{{de Ravel} et~al.,}{{de Ravel} et~al.}{2009}]{de_ravel09}
{de Ravel} L.,  et~al., 2009, \mn@doi [\aap] {10.1051/0004-6361/200810569}, \href {http://adsabs.harvard.edu/abs/2009A%26A...498..379D} {498, 379}

\bibitem[\protect\citeauthoryear{de Ravel et~al.,}{de~Ravel et~al.}{2011}]{de_ravel_zcosmos_2011}
de Ravel L.,  et~al., 2011, arXiv:1104.5470 [astro-ph]

\bibitem[\protect\citeauthoryear{{de la Calleja} \& {Fuentes}}{{de la Calleja} \& {Fuentes}}{2004}]{Calleja04}
{de la Calleja} J.,  {Fuentes} O.,  2004, \mn@doi [MNRAS] {10.1111/j.1365-2966.2004.07442.x}, \href {https://ui.adsabs.harvard.edu/abs/2004MNRAS.349...87D} {349, 87}

\bibitem[\protect\citeauthoryear{{van der Wel}, {Rix}, {Holden}, {Bell}  \& {Robaina}}{{van der Wel} et~al.}{2009}]{van_der_wel09}
{van der Wel} A.,  {Rix} H.-W.,  {Holden} B.~P.,  {Bell} E.~F.,   {Robaina} A.~R.,  2009, \mn@doi [\apjl] {10.1088/0004-637X/706/1/L120}, \href {http://adsabs.harvard.edu/abs/2009ApJ...706L.120V} {706, L120}

\bibitem[\protect\citeauthoryear{{van der Wel} et~al.,}{{van der Wel} et~al.}{2012}]{van_der_wel_12}
{van der Wel} A.,  et~al., 2012, \mn@doi [\apjs] {10.1088/0067-0049/203/2/24}, \href {http://adsabs.harvard.edu/abs/2012ApJS..203...24V} {203, 24}

\makeatother
\end{thebibliography}

\bsp	
\label{lastpage}
\end{document}